\documentclass{acm_proc_article-sp}

\usepackage[numbers,sort&compress]{natbib}
\usepackage{wrapfig}
\usepackage{graphicx}
\usepackage{tabularx}	            
\usepackage{booktabs}             

\begin{document} 

\title{Parallelization in Scientific Workflow Management Systems}


\numberofauthors{2}
\author{
\alignauthor
Marc Bux\titlenote{To whom correspondence should be addressed}\\
       \affaddr{Department of Computer Science}\\
			 \affaddr{Humboldt-Universit\"{a}t zu Berlin}\\
       \affaddr{Unter den Linden 6}\\
       \affaddr{10099 Berlin, Germany}\\
       \email{buxmarcn@informatik.hu-berlin.de}
\alignauthor
Ulf Leser\\
       \affaddr{Department of Computer Science}\\
			 \affaddr{Humboldt-Universit\"{a}t zu Berlin}\\
       \affaddr{Unter den Linden 6}\\
       \affaddr{10099 Berlin, Germany}\\
       \email{leser@informatik.hu-berlin.de}}

\maketitle

\hyphenation{KNIME SWfMS Taverna Kepler SciCumulus e-Science}


\begin{abstract}
Over the last two decades, scientific workflow management systems (SWfMS) have emerged as a means to facilitate the design, execution, and monitoring of reusable scientific data processing pipelines. At the same time, the amounts of data generated in various areas of science outpaced enhancements in computational power and storage capabilities. This is especially true for the life sciences, where new technologies increased the sequencing throughput from kilobytes to terabytes per day. This trend requires current SWfMS to adapt: Native support for parallel workflow execution must be provided to increase performance; dynamically scalable ``pay-per-use'' compute infrastructures have to be integrated to diminish hardware costs; adaptive scheduling of workflows in distributed compute environments is required to optimize resource utilization. In this survey we give an overview of parallelization techniques for SWfMS, both in theory and in their realization in concrete systems. We find that current systems leave considerable room for improvement and we propose key advancements to the landscape of SWfMS.
\end{abstract}



\section{Introduction}
\label{Introduction}

Over the last two decades, the scientific community witnessed the establishment of computation as an integral part of research beside the traditional paradigms of theory and experiment~\citep{Deelman09}. Today's scientific experiments typically involve running and refining a series of intertwined computational analysis and visualization tasks on large amounts of data. The complexity of these so-called analysis pipelines resulting in high costs for development and maintenance, the need for sharing knowledge encoded in these pipelines as well as hardware to execute them, and the need for repeatability and rigorous tracing of pipeline runs, eventuated in the emergence of \mbox{e-Science}~\citep{Hey09} and scientific workflows~\citep{Barker07}.

\begin{figure}[h]
  \centering
  \includegraphics[width=\columnwidth]{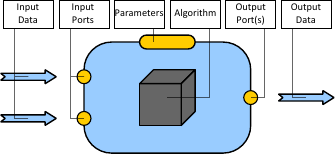}
  \caption[The structure of a single workflow task.]{The structure of a single workflow task. A task receives input data at its input ports. This data is processed according to a certain algorithm (service invocation, program call, etc.) and parameters. Generated output data is passed from the output ports to follow-up tasks for further processing.}
  \label{Node}
\end{figure}

\begin{figure*}[p]
  \centering
  \includegraphics[width=\textwidth]{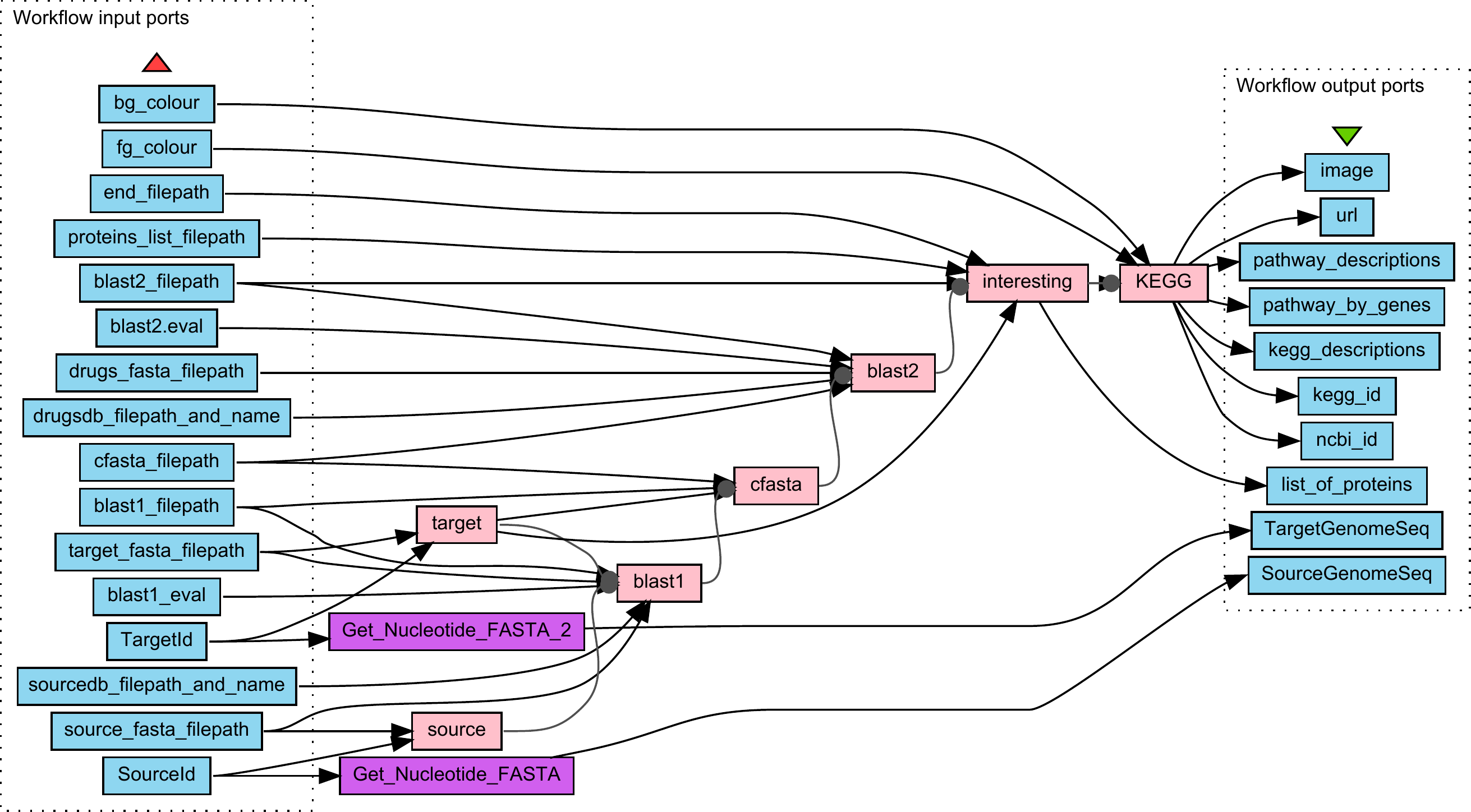}
	\caption[A Taverna workflow for finding drug targets.]{A Taverna~\citep{Oinn2004} bioinformatics workflow. Lists of proteins are retrieved and compared for pathogenic (disease) and non-pathogenic (healthy) genomes. Proteins unique to the pathogenic genome are located in KEGG pathways and investigated for the use as potential drug targets. The workflow was published in the myExperiment~\citep{Goble07} workflow repository under the id 1172. Nodes colored in light pink constitute subworkflows which have been collapsed for easier readability.}
  \label{TavernaWorkflow1172}
\end{figure*}

\begin{figure*}[p]
  \centering
  \includegraphics[width=\textwidth]{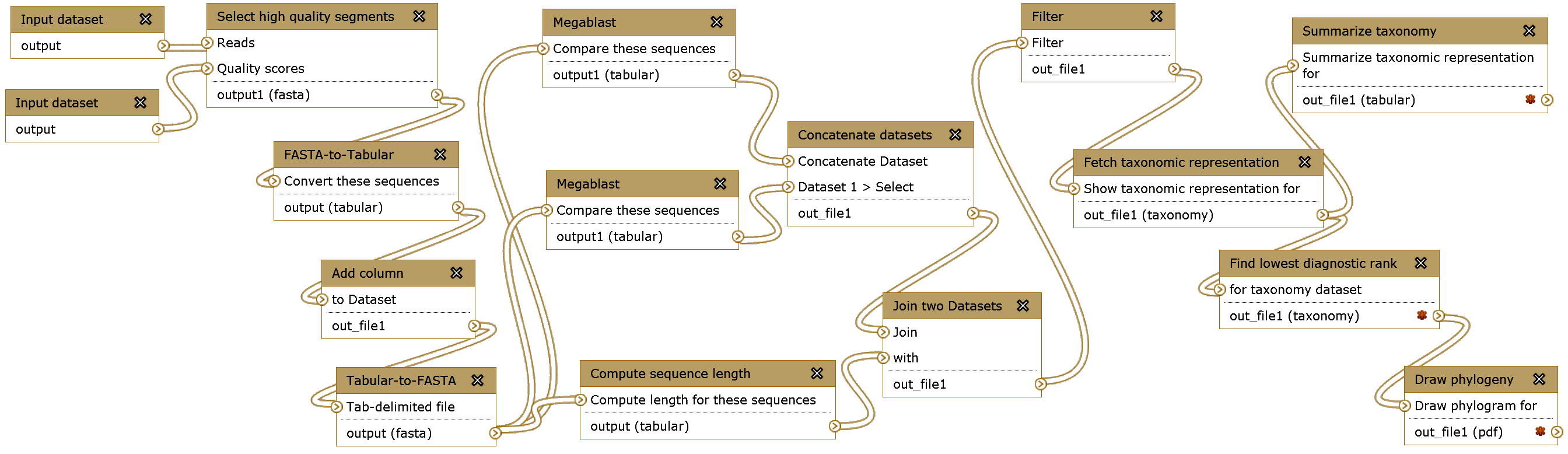}
  \caption[A Galaxy workflow for performing a metagenomic analysis.]{A Galaxy~\citep{Goecks10} workflow for performing a metagenomic analysis on next-generation sequencing data. A metagenomic analysis compares the genomes of species in one environment to the genomes of species in another environment to find environment-specific genes. The workflow was published on the public Galaxy server under the name ``metagenomic analysis''.}
  \label{GalaxyWorkflow}
\end{figure*}

Scientific workflows are compositions of sequential and concurrent data processing tasks, whose order is determined by data interdependencies \citep{Gil2007}. A task is the basic data processing component of a scientific workflow, consuming data from input files or previous tasks and producing data for follow-up tasks or output files (see Figure~\ref{Node}). A scientific workflow is usually specified in the form of a directed, acyclic graph (DAG), in which individual tasks are represented as nodes. Scientific workflows exist at different levels of abstraction: \textit{abstract}, \textit{concrete}, and \textit{physical}. An abstract workflow models data flow as a concatenation of conceptual processing steps. Assigning actual methods to abstract tasks results in a concrete workflow. If this mapping is performed automatically, it is called workflow planning \citep{Redetzki2006}. To execute a concrete workflow, input data and processing tasks have to be assigned to physical compute resources. In the context of scientific workflows, this assignment is called scheduling and results in a physical and executable workflow~\citep{Deelman2003}. Low-level batch scripts are a typical example of physical workflows. See Figures~\ref{TavernaWorkflow1172} and~\ref{GalaxyWorkflow} for two concrete bioinformatics workflows. 

\looseness=1\citet{Deelman09} constitute four phases of the workflow lifecycle: (1) the design and composition of concrete workflows; (2) the mapping of concrete workflows to the underlying physical resources (scheduling); (3) the execution of physical workflows; (4) the recording of metadata and provenance information at all stages of the workflow lifecycle. Scientific workflow management systems (SWfMS) address this lifecycle or subsets thereof by providing capabilities for modeling, executing, monitoring and storing scientific workflows. Most SWfMS operate on concrete workflows by requesting users to specify compositions of concrete processing tasks.

\looseness=1The last years have seen considerable progress in the development and deployment of SWfMS. Examples for concrete systems include Taverna~\citep{Oinn2004}, Kepler~\citep{Ludascher06}, Pegasus~\citep{Deelman05}, and KNIME~\citep{Berthold2006} (see Table~\ref{char}). A comprehensive list of universal and bioinformatics SWfMS, both commercial and from the public domain, has been assembled by \citet{Tiwari07}. To narrow the gap between domain scientists and application developers, the majority of these SWfMS involve a graphical user interface with drag and drop functionality to facilitate the composition of tasks into workflows~\citep{Cohen-Boulakia11}.

While most SWfMS provide a general-purpose framework for workflow enactment, others exhibit a certain affinity for a confined field of science and inherently provide domain-specific components. For instance, Taverna and KNIME provide built-in libraries for the fields of computational biology and chemistry, respectively. Further, some systems have been designed to cater exclusively to a specific user group or scientific domain. Frameworks like Galaxy~\citep{Goecks10}, Mobyle~\citep{Neron09} or Conveyor~\citep{Linke11} have been developed solely for applications in the life sciences. Clearly, there are trade-offs to consider between domain-specific and general-purpose approaches. While confinement to a particular field facilitates use for domain scientists and may help to promote design standards, it requires SWfMS providers to keep built-in components up to date and limits versatility and interoperability of workflow design.

\begin{table}[t]
  \caption[Most popular SWfMS.]{Most popular publicly available SWfMS}
	\label{char}
	\medskip
	\centering
  \footnotesize
	\begin{tabular}{lllll}
		\toprule
		SWfMS&languages&URL&Ref\\
		\midrule
		Taverna&Scufl, T2flow&taverna.org.uk&\citep{Oinn2004}\\
		Kepler&Ptolemy II&kepler-project.org&\citep{Ludascher06}\\
		Pegasus&DAGMan&pegasus.isi.edu&\citep{Deelman05}\\
		KNIME&&knime.org&\citep{Berthold2006}\\
		Swift&Karajan&ci.uchicago.edu/swift&\citep{Zhao07}\\
		Galaxy&&galaxy.psu.edu&\citep{Goecks10}\\
		\bottomrule
	\end{tabular}
\end{table}

In this work, we distinguish three major classes of SWfMS:

\begin{itemize}
	\item \textit{Textual workflow languages}: This category consists of low-level textual languages catered to computer-savvy users adept at using batch scripts and programming languages. Workflows are specified in the form of often complex configuration files, which are interpreted and executed by the SWfMS. The Pegasus workflow management system~\citep{Deelman05} and Swift scripting language~\citep{Zhao07} are typical examples for this group of systems.
	\item \textit{Graphical workflow systems}: SWfMS belonging to this group put a strong emphasis on ease of use. They provide a graphical user interface for workflow design and execution monitoring as well as a range of general purpose and often domain-specific task libraries. While some of these systems utilize a textual workflow language for internal representation (see Table~\ref{char}), this language is not intended and designed to be accessed by the user. Examples include the Taverna~\citep{Oinn2004} and Kepler~\citep{Ludascher06} workflow systems.
	\item \textit{Domain-specific web portals}: This category comprises online portals, where scientists can design, execute, and share workflows within a certain domain. In most cases no software has to be installed locally. Workflows are composed exclusively from built-in components in a web browser and are executed on a public (or private) server. For instance, Galaxy~\citep{Goecks10} and Mobyle~\citep{Neron09} can be considered as an enactment portal for research in the life sciences. \end{itemize}

\looseness=1This categorization was inspired by the work of \citet{Romano08}, who proposed to subdivide SWfMS in the life sciences into \textit{software libraries}, \textit{standalone systems}, \textit{client/server systems}, and \textit{enactment portals}. We utilize a slightly different categorization, as we found most standalone SWfMS to provide a client/server-based implementation as well and hence found it difficult to differentiate between the two. Also, we don't consider mere software libraries to qualify as SWfMS.

\begin{figure}[p]
\begin{center}
  \includegraphics[width=\columnwidth]{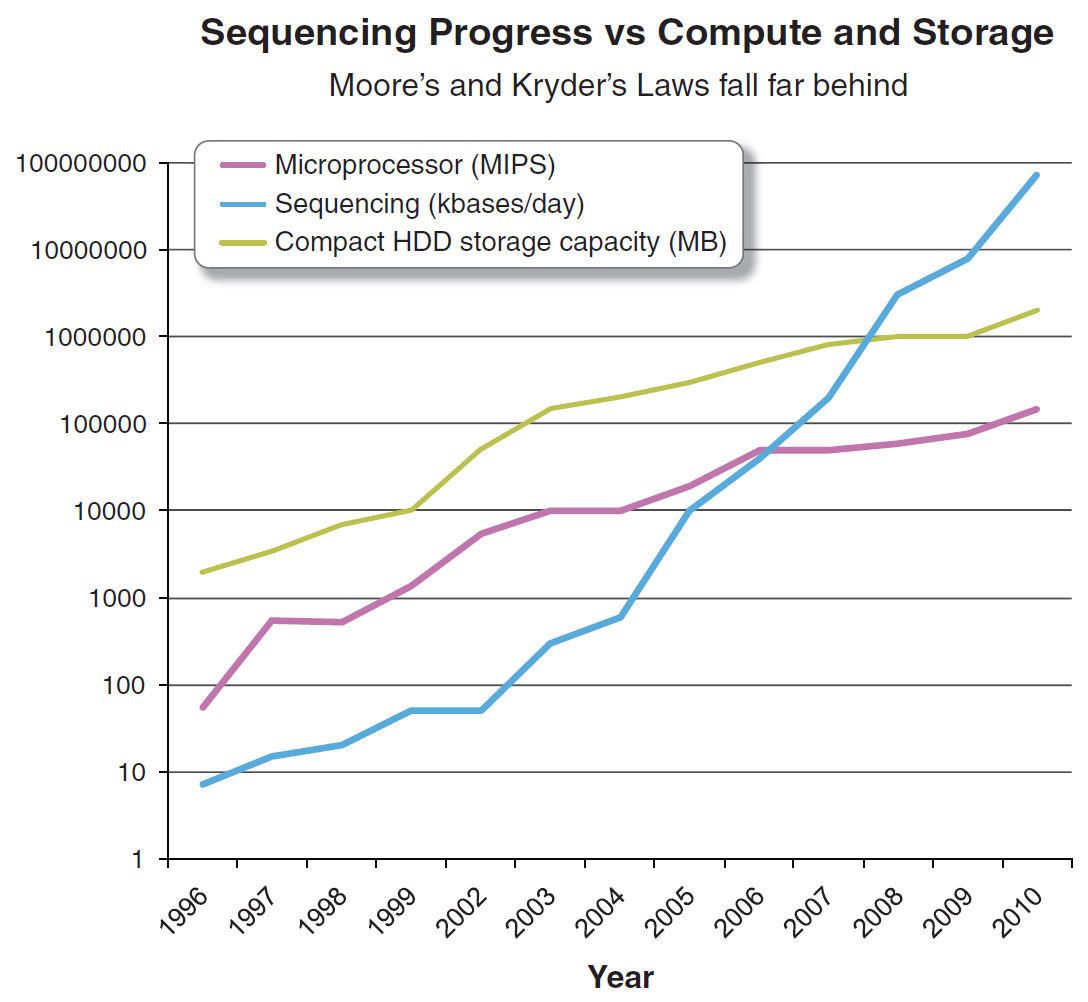}
\end{center}
  \caption{Development of processor speed, HDD storage capacity and genomic data produced (in 1000 nucleic acids per day). Image taken from~\citep{Kahn2011}.}
  \label{capacity}
\end{figure}

The wide range of SWfMS outlined above provides solutions for analysis pipelines of researchers from various domains. However, due to ever-increasing amounts of data across all fields of science, the computational effort required to execute a given scientific workflow is becoming more and more critical. In fact, the magnitude of data produced in many scientific domains has risen at exponential rates and often outpaced advances in storage capacity, network bandwidth, and processing power. In bioinformatics, for instance, recent years brought a new generation of devices, which produce genomic data at unprecedented scale~\citep{Pennisi2011}. The generation of genomic data was found to double every nine months -- at a pace much faster than computing power and storage capacity (see Figure~\ref{capacity}).

\begin{figure}[p]
\begin{center}
  \includegraphics[width=\columnwidth]{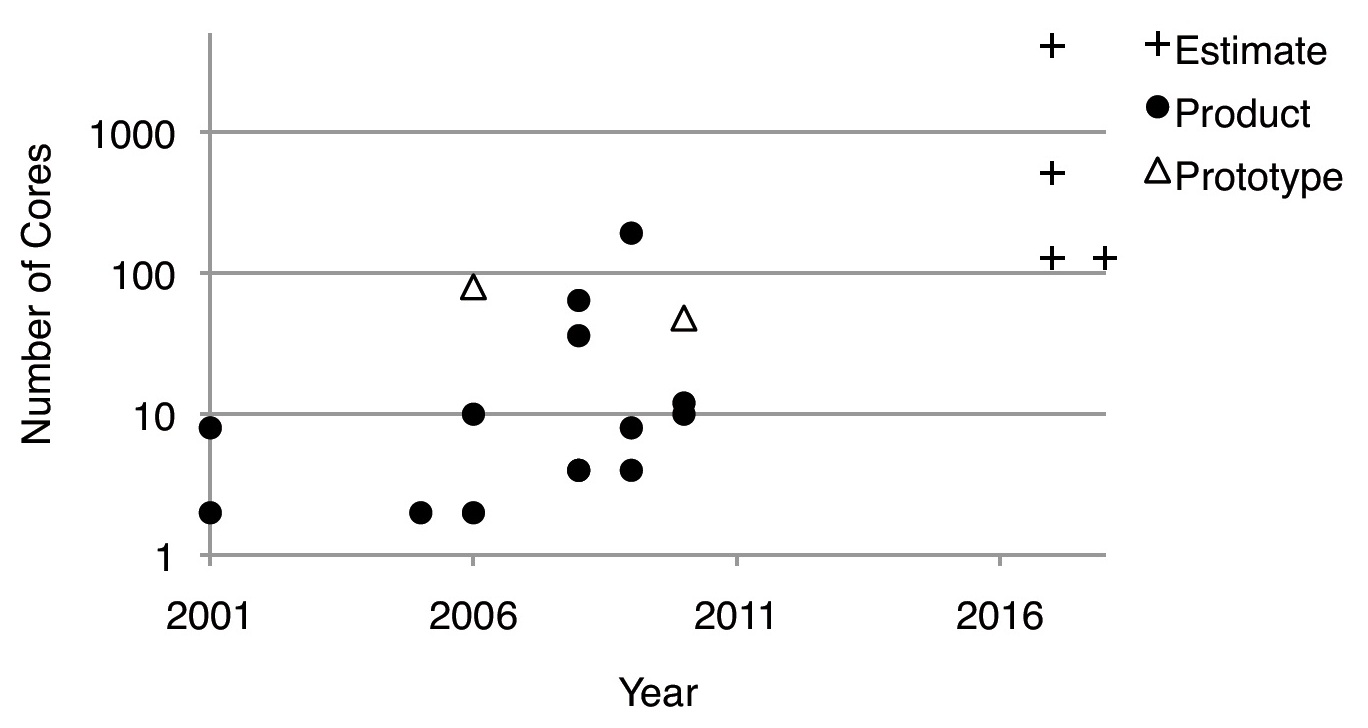}
\end{center}
  \caption[Chip development over the last decades.]{Development of the number of cores per chip over the last decades. Image taken from \citep{Plattner11}.}
  \label{Hasso}
\end{figure}

\begin{figure}[p]
\begin{center}
  \includegraphics[width=\columnwidth]{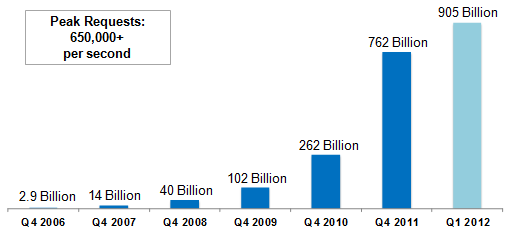}
\end{center}
  \caption{Total number of objects stored in the Amazon Simple Storage Service (Amazon S3) since its introduction in 2006. Image has been published on the Amazon Web Services Blog in April 2012\protect\footnotemark.}
  \label{s3growth}
\end{figure}

Besides algorithmic advances, the canonical way to deal with increasing data volumes is parallelization. This is true for all areas of computer science and is reflected in the development of parallel execution of threads on single chips as well as on infrastructures which combine multiple machines to clusters, grids, and clouds. Figure~\ref{Hasso} displays the development of number of cores per CPU, as observed over the last decades. It showcases the trend towards multicore architectures in chip design. Figure~\ref{s3growth} illustrates the number of objects stored in Amazon's Simple Storage Service (Amazon S3). The exponential growth of the largest cloud provider's data storage solution mirrors the trend of parallel computation environments such as compute clouds increasing in size and popularity.

\setcounter{footnote}{-1}

While several SWfMS like DAGMan and Pegasus have been designed with parallel computation on shared resources in mind, they rarely provide multicore support and are generally difficult to set up and utilize by domain scientists. As a result, there has been little uptake in the scientific community and installations of these systems on actual data centers are few and far between~\citep{Cohen-Boulakia11}. More recent systems like Taverna put a stronger emphasis on usability, yet provide only limited means towards parallelization and utilization of distributed compute resources. This survey explores the gap between inherent parallelization and ease of use in current SWfMS.

\footnotetext{http://tinyurl.com/6uh8n24}

To this end, we present an overview of parallelization techniques for SWfMS. Workflow systems differ strongly in their support for parallelism and we believe that there is an urgent need for a comparative survey focusing on this aspect. We hope that this will serve as an entry point for both domain scientists with data-intensive workflows at hand and SWfMS researchers with an interest in parallel computing. The three main contributions of this survey are:

\begin{itemize}
	\item A taxonomy for approaches towards parallelism in scientific workflow management systems. This taxonomy focuses on concepts which are treated differently by many SWfMS and can arguably be improved upon. At the same time, the taxonomy omits aspects which are either not related to parallelization or don't serve as distinctive features since they are implemented identically in most SWfMS.
	\item A comparative overview of parallelization techniques and computational infrastructures supported by current SWfMS. We believe that such a contrasting juxtaposition can serve both as a reference for researchers and as a starting point for scientific workflow users with large amounts of scientific data at hand.
	\item An outline of current trends along with a discussion of future research questions that could leverage scientific workflows as the standard model of computation for parallel execution of computatianally intensive intensive analysis pipelines.
\end{itemize}

\looseness=1To the best of our knowledge, an overview of parallelization techniques for SWfMS in the era of cloud computing has not been published yet. Perhaps the most similar work to ours has been conducted in 2006 by \citet{Yu2006}, who presented a comprehensive taxonomy of SWfMS for grid computing. Their taxonomy is comprised of close to a hundred terms and is strongly connected to execution on a grid infrastructure. In this survey, we chose to employ a more general, light-weight taxonomy and discuss current SWfMS running on a wider range of computational infrastructures.

The rest of the paper is organized in the following way. Fundamental concepts of parallelism in scientific workflows are outlined in Section~\ref{Concepts}. This encompasses types of parallelism, parallel computation infrastructures, and scheduling techniques. Concrete realizations of parallelism in current scientific workflow management systems are described in Section~\ref{Systems}. We then discuss arising research questions and summarize in Section~\ref{Discussion}.

\section{Aspects of parallel scientific workflow execution}
\label{Concepts}
\begin{figure}[h]
  \centering
  \includegraphics[width=8cm]{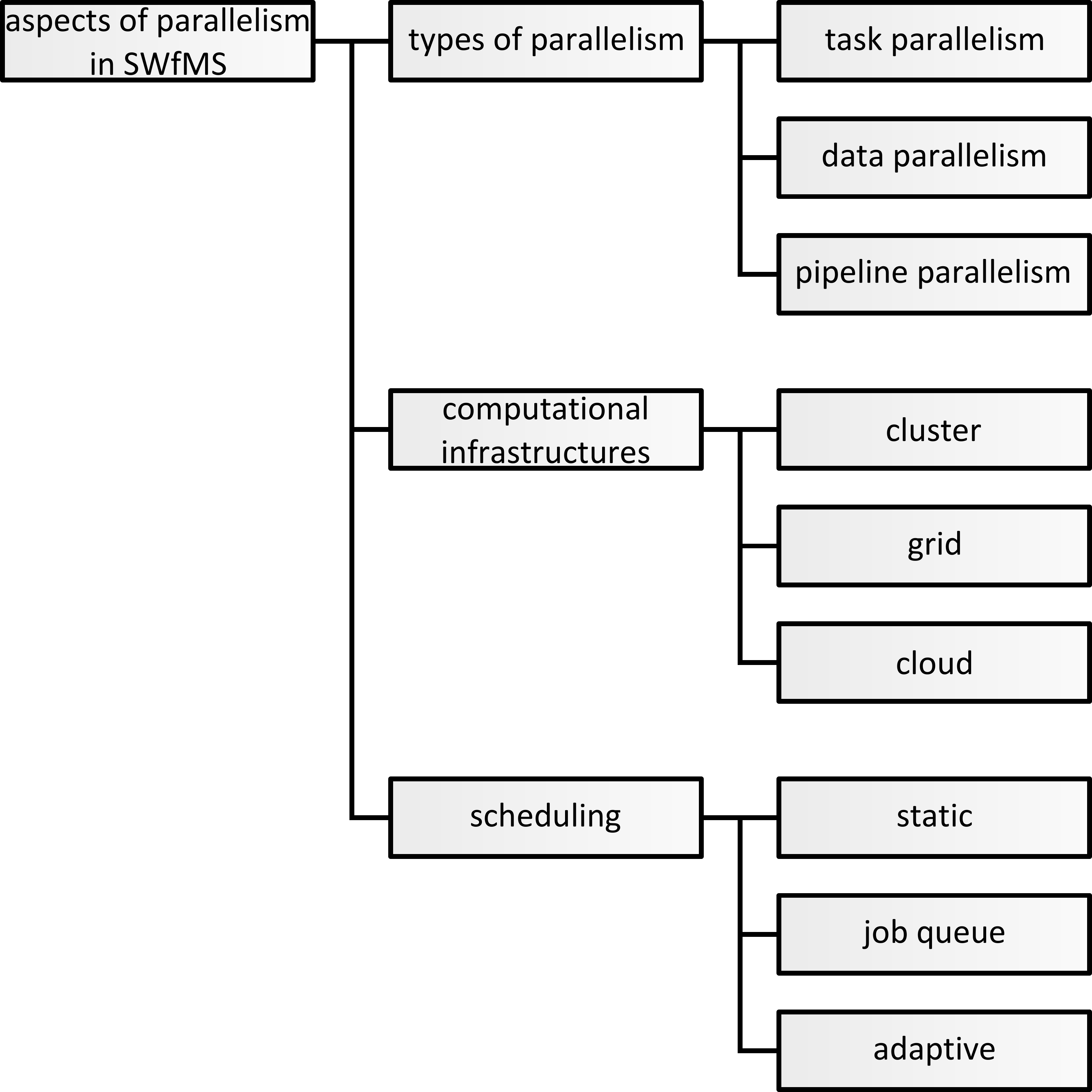}
  \caption[Taxonomy on parallelization of scientific workflows.]{A taxonomy on the most important aspects of parallelization in scientific workflows.}
  \label{Taxonomy}
\end{figure}

Scientific workflow systems vary greatly in their means to accomplish parallel computation. In this section, we outline and contrast fundamental strategies towards parallelization. We differentiate between basic types of parallelism, parallel computation infrastructures, and different scheduling policies. We will later adopt these categories to compare different realizations of parallelism in concrete SWfMS. A graphical overview of our taxonomy is given in Figure~\ref{Taxonomy}.

We highlight the introduced concepts on the basis of an exemplary workflow from the field of bioinformatics. To this end, we use a workflow described by~\citet{Li2008}. Genomic sequencing aims at revealing the ordered sequence of nucleic acids (DNA) in a given sample, such as a human chromosome. Until today, this can only be achieved by splitting the DNA sequence into many short overlapping pieces, which are called reads.

\begin{figure*}[htbp]
  \centering
  \includegraphics[width=\textwidth]{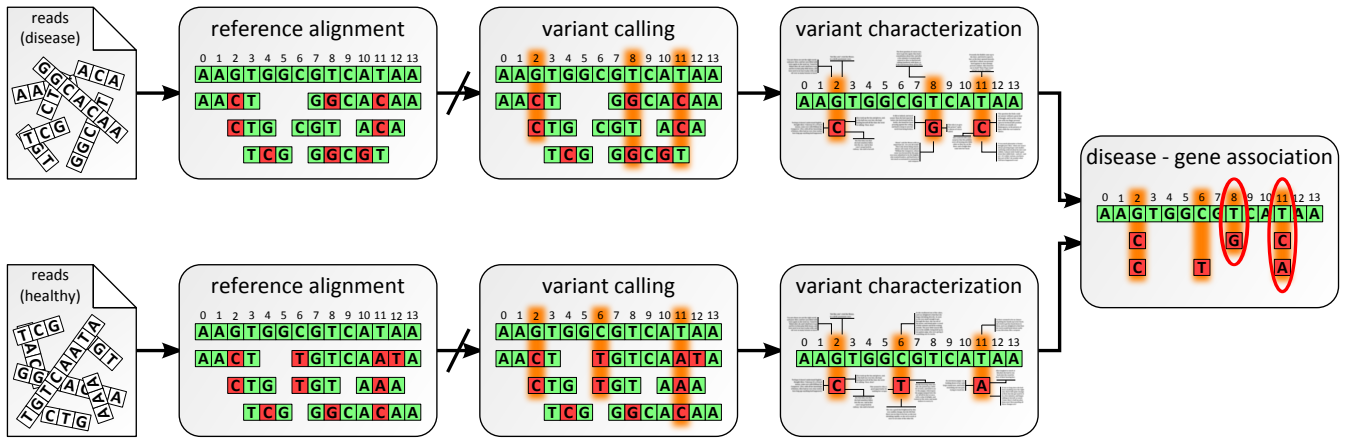}
  \caption[An exemplary workflow from the field of bioinformatics.]{A bioinformatics workflow which processes genomic sequencing data~\citep{Li2008}. In reference alignment, two sets of DNA reads -- from a disease and healthy sample respectively -- are mapped onto an established (and different) reference genome. Alignments are investigated for variants: mismatching nucleic acids, which might be indicative of mutations. Variant sets are then characterized and compared between the two samples. Mutations specific to the disease sample might be functionally associated to the disease. The crossed out data dependency link between the reference alignment and variant calling tasks indicates variant calling being a \textit{pipeline blocker}, i.e.,~it can't commence until reference alignment has finished processing all of its data.}
	\label{AlignmentWorkflow}
\end{figure*}

The workflow in Figure~\ref{AlignmentWorkflow} processes genomic sequencing data in the form of such reads. It requires as input two sets of reads from different environments, such as healthy and pathogenic (disease) tissue. In the first step, which is referred to as reference alignment, these reads are mapped to a different, much larger reference genome. The aligned reads are then compared to the reference genome in detail to detect variants, which serve as indicators for mutations. Characteristics of detected variants such as rarity are obtained from external databases like dbSNP\footnote{http://www.ncbi.nlm.nih.gov/projects/SNP/}. In the final step, the obtained sets of variants are compared between healthy and pathogenic tissue. Mutations specific to the pathogenic genotype might be related to the disease. The associated gene could therefore qualify as a drug target~\citep{Li2008}.

\looseness=1The workflow displayed in Figure~\ref{AlignmentWorkflow} is an \textit{abstract} workflow. The data processing steps reference alignment, variant calling, and disease-gene association, constitute abstract concepts for which concrete algorithms have not been selected yet. In the following section, we will use this workflow as a reference to showcase different parallelization techniques.

\subsection{Types of parallelism}
\label{Parallelization}
\looseness=1A scientific workflow describes the processing of data by a set of tasks. Parallelization addresses the question of how to distribute the workload associated with a given workflow on several compute nodes. There are different means to accomplish parallelization, all of which involve subdividing either the set of workflow tasks or the input data (or both). An important concept in this context is the degree of parallelism, which we define as the number of concurrently running machines or threads at any given time and which can vary for a given workflow depending on the utlized type of parallelism.

In this section, we distinguish three major types of parallelization~\citep{Gordon06}: task, data, and pipeline parallelism. We describe the characteristics of and relations between these three approaches to parallelism. We discuss prerequisites that have to be met, problems than can occur, and the degree of parallelism that can be achieved. All concepts we discuss apply equally well to the level of multiple machines or multiple threads on a single machine. For ease of exposition, in the following discussion we shall focus on the former case.

\subsubsection*{Task parallelism}
\begin{figure*}[htbp]
  \centering
  \includegraphics[width=\textwidth]{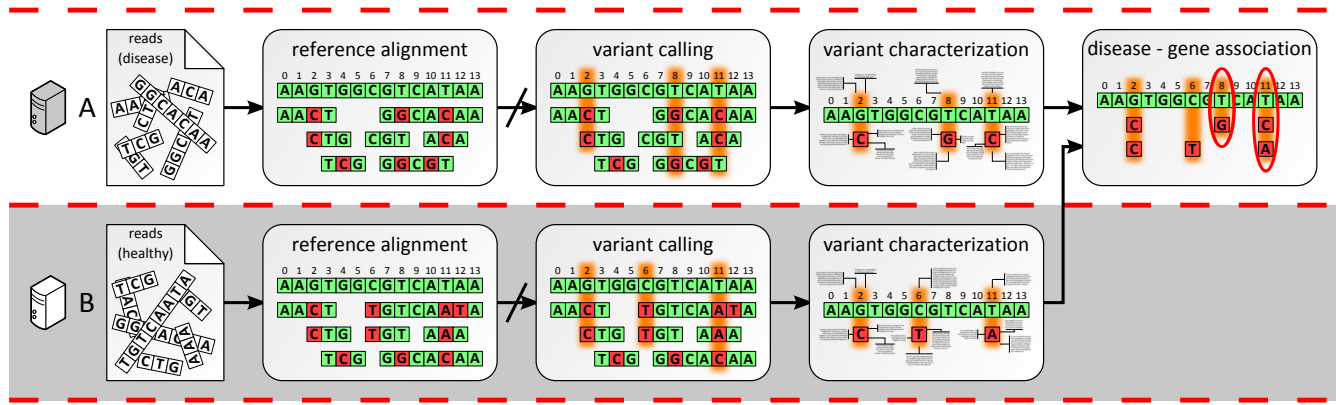}
  \caption[Task parallelism in the example of a DNA sequence alignment workflow.]{Task parallelism illustrated by the bioinformatics workflow from Figure~\ref{AlignmentWorkflow}. The separated horizontal lanes correspond to different compute resources processing different tasks in the course of time.	The reference alignment, variant calling and variant characterization tasks can be run independently of another. In this instance they are run task parallel on the compute resources A and B, resulting in a degree of parallelism of two.}
  \label{TypesOfParallelismHorizontalTask}
\end{figure*}

Task parallelism is achieved when the tasks composing a workflow are distributed over several independent compute nodes. It is only applicable to tasks located on parallel branches of the workflow. Data dependencies and the overall number of components within the workflow graph strongly limit opportunities for task parallelism. The maximum number of parallel tasks can be computed easily in advance by analyzing the workflow structure. However, choosing the right scheduling strategy is not trivial, as parallel tasks might exhibit varying runtimes. Differences in task runtimes are also the reason for junctions in the data flow being difficult to handle: tasks requiring input data from several concurrent parent tasks have to wait until all parent tasks have finished execution. Buffering of intermediate results or synchronization of task execution is therefore required for task parallelism.

A major advantage of task parallelism is that existing workflows need not be adjusted for tasks to be run concurrently, because the graph structure of the workflow alone provides sufficient information. This makes task parallelism fairly easy to implement. With respect to the example introduced in Figure~\ref{AlignmentWorkflow}, the tasks of reference alignment, variant calling, and variant characterization are suitable for parallel execution, as they lie on parallel branches of the workflow and don't have any data dependencies in between (see Figure~\ref{TypesOfParallelismHorizontalTask}). Evidently, the achievable degree of parallelism is fairly limited for workflows which are mostly comprised of sequential tasks.

\subsubsection*{Data parallelism}
\begin{figure*}[htbp]
  \centering
  \includegraphics[width=\textwidth]{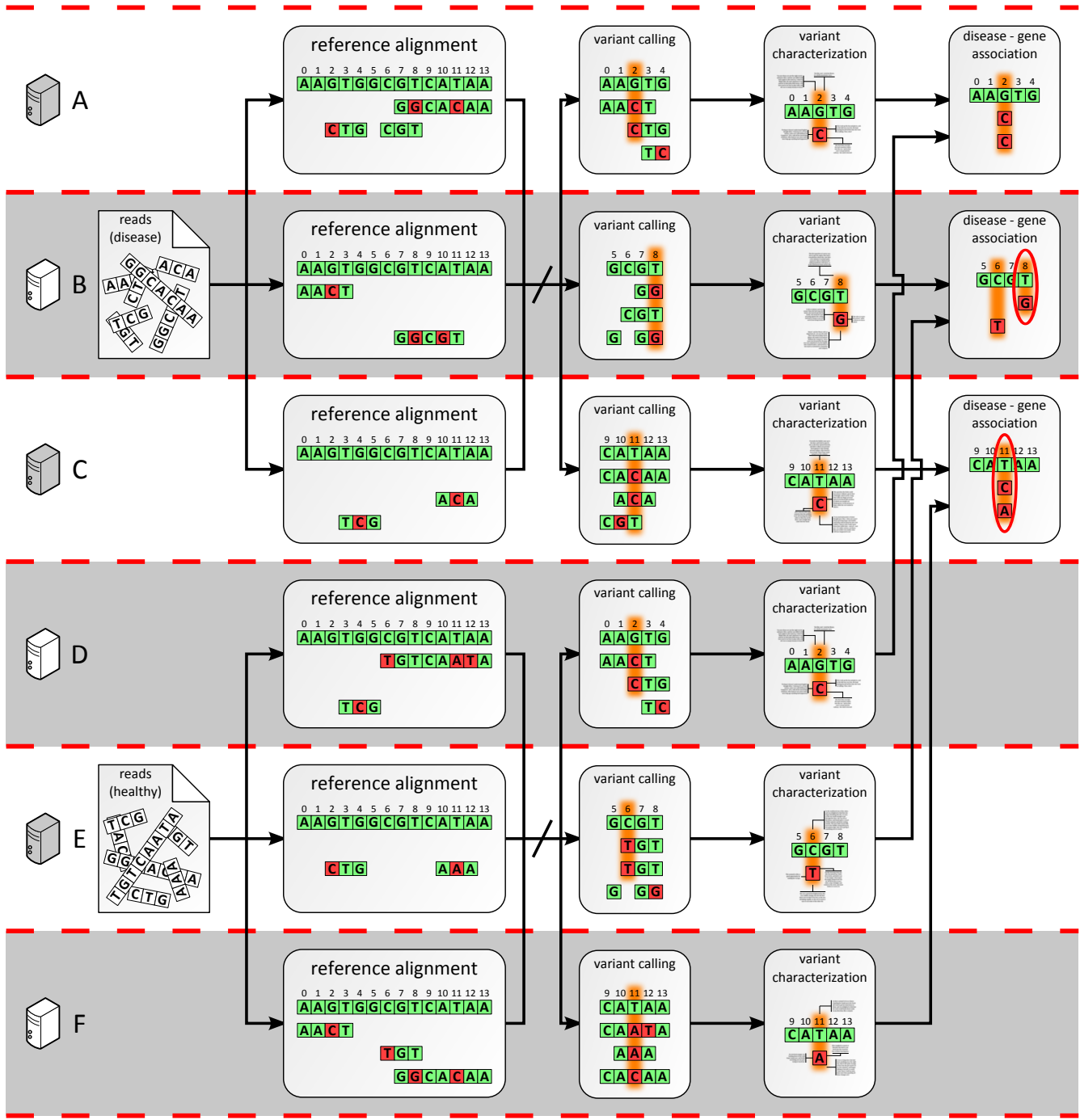}
  \caption[Data parallelism in the example of a DNA sequence alignment workflow.]{A possible implementation of data parallelism in the workflow introduced in Figure~\ref{AlignmentWorkflow}. Again, horizontal lanes correspond to different compute resources. The alignment task is embarrassingly parallel and can be replicated for different fragments of input data. Since variant calling is a pipeline blocker, all the alignments have to be merged before further processing can commence. Subsequent to this merger, variant calling, variant characterization, and disease--gene association are performed data parallel on fragments of the alignment. Since these tasks are not embarrassingly parallel, additional effort has to be put into how to split and merge input and output files.}
  \label{TypesOfParallelismHorizontalData}
\end{figure*}

In data parallelism, input or intermediate data is split into distinct chunks, each of which is processed on a different compute node.  This means that the workflow -- or a part thereof -- is replicated on each compute node for a different fragment of the data. Depending on the granularity of data (i.e.,~how many chunks the data can be split into), very high degrees of parallelism are achievable.

Data parallelism is only feasible if data can be split into independent chunks. Most suitable for data parallelism are so-called \textit{embarrassingly parallel} problems in which data items can be processed independently from each other. In the bioinformatics workflow introduced in Figure~\ref{AlignmentWorkflow}, the reference alignment task is an embarrassingly parallel problem because every read can be mapped to the reference independently of all others. In contrast, partitioning the data in variant calling is much more complex since all reads overlapping a position in the reference have to be considered jointly to tell true variants from noise in single reads. See Figure~\ref{TypesOfParallelismHorizontalData} for a possible implementation of data parallelism in the bioinformatics workflow introduced earlier.

In general, we distinguish three techniques for achieving data parallelism: (1) data can be manually split and distributed among several replicate processing tasks, effectively exploiting concepts of task parallelism; (2) scientific workflows can be specified \textit{ad hoc} in a language inherently supporting data parallelism, such as languages based on the MapReduce programming paradigm~\citep{Dean08}; (3) computationally demanding tasks within a workflow can be annotated \textit{post hoc} with metadata describing how data is to be processed in parallel. All of these methods require either additional expenditures from the user or extended functionality from the SWfMS to exploit data parallelism. Also, the techniques vary strongly in effort to set up and potential for parallel execution.

\subsubsection*{Pipeline parallelism}
\begin{figure}[htbp]
  \centering
  \includegraphics[width=\columnwidth]{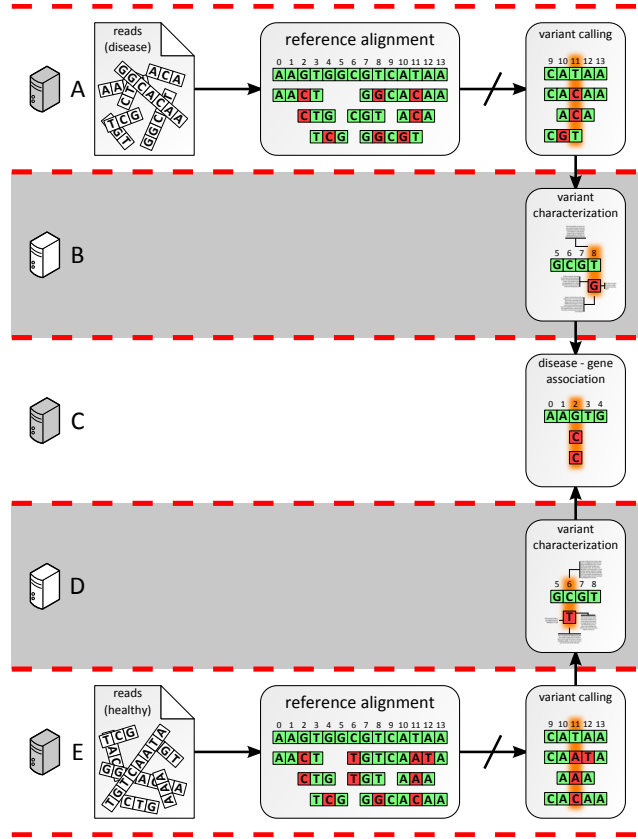}
  \caption[Pipeline parallelism on the example of a DNA sequence alignment workflow.]{Pipeline parallel execution of the workflow from Figure~\ref{AlignmentWorkflow}. Again, horizontal lanes correspond to different compute resources. There is a synchronization point between reference alignment and variant calling, since variant calling is a pipeline blocker. Subsequent to reference alignment, different chunks of the alignment data are located and processed at different stages of the processing pipeline. While the first fragment of data already undergoes disease--gene association, variants in the second fragment are characterized, and the third fragment is still investigated for variants. All of these tasks are executed in parallel on different compute resources. If the tasks vary in their computational cost, this assignment of resources certainly is not optimal. Additional resources could be assigned to slower tasks, allowing for a more fine-grained balancing of workload among resources.}
  \label{TypesOfParallelismHorizontalPipeline}
\end{figure}

In pipeline parallelism, sequential steps of data processing are executed simultaneously on different parts of the input data. Thus, partial output data produced by workflow tasks is passed to follow-up tasks for immediate consumption -- similar to workers on an assembly line. Pipeline parallelism shares characteristics with task and data parallelism and can be considered a subset of both:

\begin{itemize}
	\item Similar to task parallelism, the tasks composing a scientific workflow are distributed over independent compute nodes. However, this distribution is not limited to tasks on parallel branches of the workflow, resulting in a potentially higher degree of parallelism.
	\item As with data parallelism, input data is fragmented and processed independently on different compute nodes. However, in contrast to data parallelism, the chronological order in which fragments of data are processed as well as the assignment of tasks to compute nodes are determined in advance and restricted by the concept of the pipeline.
	\item At its finest granularity, pipeline parallelism is closely related to streaming~\citep{Gordon06}.
\end{itemize}

One common problem especially in scientific applications emerges if tasks composing a sequential workflow vary in their computational cost. In order to execute sequential tasks with different runtime in parallel, advanced scheduling and buffering of intermediate results is essential. Another problem arises if a task requires input data to be present as a whole before it is able to start execution (a so-called \textit{pipeline blocker}). A prominent example of such a task is sorting.

Figure~\ref{TypesOfParallelismHorizontalPipeline} illustrates pipeline parallel execution of the workflow introduced in Figure~\ref{AlignmentWorkflow}. After some execution time has passed, different fragments of data are located at different stages of the processing pipeline, processed by different compute resources. In this example, all tasks are assigned their own compute node. This certainly isn't the optimal assignment if the tasks vary strongly in their computational cost.

\subsubsection*{Hybrid schemes}
\begin{figure}[t]
  \centering
  \includegraphics[width=\columnwidth]{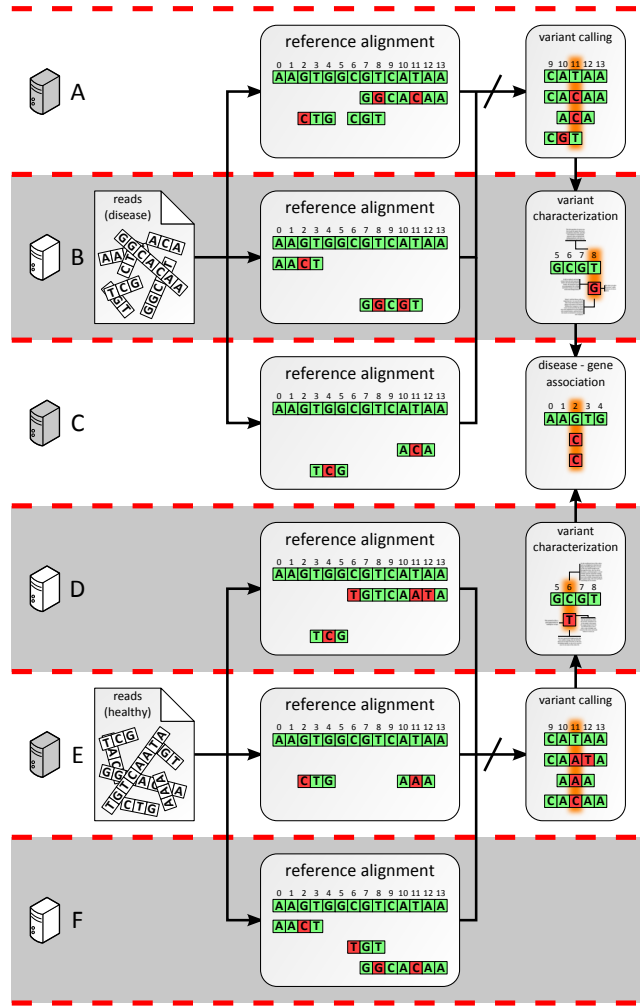}
  \caption[Pipeline parallelism on the example of a DNA sequence alignment workflow.]{Implementation of the workflow from Figure~\ref{AlignmentWorkflow} utilizing task, data and pipeline parallelism. Again, horizontal lanes correspond to different compute resources. While reference alignment is performed on concurrently on different subsets of input data, the following tasks are executed in the form of a data processing pipeline.}
  \label{AlignmentWorkflowHybridVertical}
\end{figure}

Task, data, and pipeline parallelism are not antagonistic by any means and can be combined for increased effect. In the case of the bioinformatics workflow introduced in Figure~\ref{AlignmentWorkflow}, reference alignment can be performed for different fragments of input data on distinct compute resources. The subsequent tasks of variant calling and characterization as well as disease--gene association can then be realized as a data processing pipeline. This realization of the workflow therefore utilizes all three types of parallelism (see Figure~\ref{AlignmentWorkflowHybridVertical}).

All of the introduced types of parallelism have the aim to split the workload associated with a scientific workflow, yet they differ in the means by which they accomplish this goal. Clearly, the abundance of parallelization techniques complicates the act of scientific workflow scheduling. Approaches towards scheduling of scientific workflows will be outlined in Section~\ref{Scheduling}.

\subsection{Computational infrastructures for parallel processing}
\label{Cloud}
Realizations of parallel workflow execution are usually designed to run in a particular computation environment. Generally, one can differentiate between three settings: local cluster, compute grid, and compute cloud. In this section, we give a short summary of these compute infrastructures, which is based on the definitions in~\citep{Foster2002,Mell11} and the work of~\citet{Foster2008}.

We define a compute cluster as a set of tightly connected computers which operate as a single system. Ever-increasing numbers of cores per CPU and CPUs per cluster (see Figure~\ref{Hasso}) have led to a strong potential for parallelization. However, up-front acquisition costs are fairly high, which is problematic if compute resources are only required when new experimental data is available for analysis. For instance, bioinformatics workflows such as the one illustrated in Figure~\ref{AlignmentWorkflow}, process genomic data, which is typically generated infrequently yet requires substantial computational effort to process.

The sporadic need for the processing capabilities of large clusters ultimately resulted in increased efforts of resource sharing by the scientific communities. In the early 1990s, grid computing was promoted as a new paradigm of distributed computing in which compute resources could be heterogeneous and geographically far away from the client. The idea was to connect compute resources of different proprietors in order to solve computationally demanding problems without the need for a supercomputer. Several SWfMS, such as Pegasus~\citep{Deelman05} or Condor DAGMan~\citep{Litzkow88,Couvares2007}, have been developed to utilize grid resources for parallel execution of computationally intensive workflows.

Cloud computing describes a more recently established form of distributed computing, which provides rentable compute and storage resources on-demand and over the Internet~\citep{Mell11}. The ``pay-per-use'' cost model of commercial cloud providers charges users by the hour and is therefore especially interesting in environments where data analysis is infrequent yet computationally intensive. Resources are generally made available at different levels of abstraction, namely infrastructure as a service (IaaS, e.g.,~Amazon Web Services\footnote{http://aws.amazon.com}), platform as a service (PaaS, e.g.,~Microsoft Windows Azure\footnote{http://www.windowsazure.com}, Google App Engine\footnote{https://developers.google.com/appengine}), and software as a service (SaaS, e.g., Google Apps\footnote{http://www.google.com/apps}, Galaxy~\citep{Goecks10})~\citep{Foster2008}.

In SaaS, concrete applications are hosted in cloud infrastructure and accessed from the client via specific interfaces. For instance, Galaxy~\citep{Goecks10} and Mobyle~\citep{Neron09} are two SWfMS for the life sciences, which can be accessed by users from a web browser and which execute workflows on shared computational infrastructure. In PaaS and IaaS, resources are provisioned in the form of virtual machines including a custom operating system and an ephemeral disk, which allows storage of intermediate results for as long as the virtual machine is leased. The fundamental difference between PaaS and IaaS lies in the fact that in IaaS the user has more control over the runtime environment and middleware, which in PaaS are managed by the cloud provider.

Many SWfMS have been extended to allow the utilization of IaaS and PaaS cloud resources (e.g.,~Pegasus~\citep{Hoffa08,Juve09,Juve11}, Swift~\citep{Zhao07}), while others especially in the life sciences have been developed particularly with the cloud environment in mind (e.g.,~\citep{Angiuoli11}, \citep{Wu11}). Elasticity, which denotes the possibility to provision additional resources at runtime, constitutes a key benefit of cloud computing, yet has not been utilized to full capacity by SWfMS.

Despite commercial cloud vendors providing guarantees with regards to processor clock speed and memory capacity, the actual performance of rented virtual machines varies greatly depending on the configuration of underlying hardware and utilization of shared resources by other users. In Amazon EC2, \citet{Dejun2010} observed response times of CPU- and I/O-intensive web applications to vary by a factor of four and two respectively. \citet{Jackson2010} found network communication between virtual machines to vary by a factor of up to 1.7 due to sharing of network resources. \citet{Zaharia2008} reported I/O performance to vary by a factor of up to 2.7, depending on how many virtual machines performed I/O operations on the same physical hardware. Apparently, compute clouds are far more heterogeneous and dynamic than commonly perceived. For SWfMS, this finding translates into an elevated importance of adaptivity in scheduling, as outlined in the next section.

While it is difficult to conduct a comprehensive comparison of scientific workflow execution on cluster, grid, and cloud infrastructures, \citet{Hoffa08} contrasted execution of the Montage workflow~\citep{Berriman2004} in Pegasus on a local machine as well as a remote grid and cloud infrastructure with up to four utilized CPUs of comparable performance. Montage consists of a very large number of tasks with a runtime on the order of a few seconds. In the grid and cloud environments, \citeauthor{Hoffa08} observed substantial wide-area data transfer and delays in instantiation of large amounts of short tasks, leading to performance degredation.

Clearly, the execution of scientific workflows presents different challenges depending on the underlying computational infrastructure. Since computational resources in clusters are tightly coupled, data locality is less of an issue compared to distributed environments, such as grids and clouds. Compute clusters also provide a more homogeneous environment in terms of CPU performance and latency / bandwidth between compute nodes. However, scalability of compute clusters is limited and up-front investment costs are arguably high.

\subsection{Workflow scheduling}
\label{Scheduling}
Scheduling a scientific workflow involves mapping concrete tasks to the available physical resources~\citep{Mandal2005}.
This usually involves optimizing a cost function which can incorporate estimates for task runtimes and compute resource performance. More sophisticated schedulers may also include network data transfer in their runtime assessment (e.g.,~\citep{Topcuoglu2002}). While not examined in this work, in certain applications it might be preferable to optimize monetary cost investment or maximize data security, as opposed to minimizing workflow execution time \citep{Yu2006}.

Since all possible placements of tasks on compute resources have to be considered, scheduling scientific workflows is NP-complete in the number of workflow tasks~\citep{Fernandez-Baca1989} (by reduction from Minimum Multiprocessor Scheduling~\citep{Garey1979}). Scheduling is therefore typically approached heuristically, for instance by restricting the scheduling perimeter (i.e.,~the maximum number of tasks considered for scheduling at the same time). The processing of large data volumes along with the existence of data dependencies between tasks separate scientific workflow scheduling from traditional scheduling in operating system design~\citep{Tanenbaum2006}.

For the task of workflow scheduling, it is highly beneficial to have estimates on the runtimes of individual components, the performance of different compute nodes and the data transfer rates between compute nodes. The accuracy of task runtime estimates is known to largely affect scheduling and overall workflow completion time \citep{Blythe2005,Morton2010}. Approaches of runtime estimation can be separated into three major groups \citep{Deelman05}:

\begin{itemize}
	\item \textit{Empirical models} (e.g.,~\citep{Yang07}) treat tasks as black boxes and model their behavior based on past performance. Since most scientific workflow management systems provide native support for capturing provenance data, performance prediction based on previous runs appears to be a promising technique.
	\item \textit{Analytical models} (e.g.,~\citep{Taylor2002}) employ historical performance data, yet attempt to describe the mechanics underlying a task as a composition of mathematical functions.
	\item \textit{Simulation techniques} (e.g.,~\citep{Mandal2005}) such as sampling require a small subset of actual (or artificial) input data to be distributed among available resources along with necessary executables. The time required for task execution and data transfer is captured and serves as an initial assessment of the compute resources' performance.
\end{itemize}

In the context of scientific workflow scheduling, performance instability on possibly shared computational infrastructure as well as dynamically changing sets of available compute resources can be problematic if schedules are generated in advance or schedulers are oblivious to changes in the computational infrastructure. Adaptive scheduling denotes the ability to adjust workflow execution to a dynamically changing compute infrastructure at runtime with the aim of minimizing time to workflow completion. One can differentiate between several levels of adaptivity in scientific workflow scheduling~\citep{Yu2006}: static, job queue, and adaptive scheduling.

In the following, we introduce these concepts using the example of the bioinformatics workflow introduced earlier and executed data parallel, as illustrated in Figure~\ref{TypesOfParallelismHorizontalData}. Figure~\ref{AlignmentWorkflowDataScheduling} contains the same workflow along with arbitrary numbers for estimated and actual runtimes of every task category (reference alignment, variant calling, variant characterization, and disease--gene association) on each out of three available compute resources A, B, and C. Different underlying hardware and network infrastructures result in the compute resources exhibiting a different runtime behavior when assigned certain tasks. For instance, compute resource B has a comparably short runtime for variant characterization (possibly due to a favorable network connection), whereas it exhibits lackluster performance for variant calling.

\begin{figure*}[htbp]
  \centering
  \includegraphics[width=\textwidth]{AlignmentWorkflowDataScheduling.pdf}
  \caption[...]{The bioinformatics workflow introduced in Figure~\ref{AlignmentWorkflow}, executed data parallel as illustrated in Figure~\ref{TypesOfParallelismHorizontalData}. Estimated and actual runtimes (in arbitrary units of time) are given for each category of tasks (reference alignment, variant calling, variant characterization, and disease--gene association) on each of three available compute resources A, B, and C. Depending on the underlying hardware configuration (CPU, I/O throughput, network speed, etc.) compute resources have a different affinity towards certain tasks. In this example, the compute resources are assumed to exhibit constant performance, though in a real-world scenario this might not be the case especially if resources are shared between users. Note that data movement is not taken into account in this example.}
  \label{AlignmentWorkflowDataScheduling}
\end{figure*}

\subsubsection*{Static scheduling}
\begin{figure*}[htbp]
  \centering
  \includegraphics[width=\textwidth]{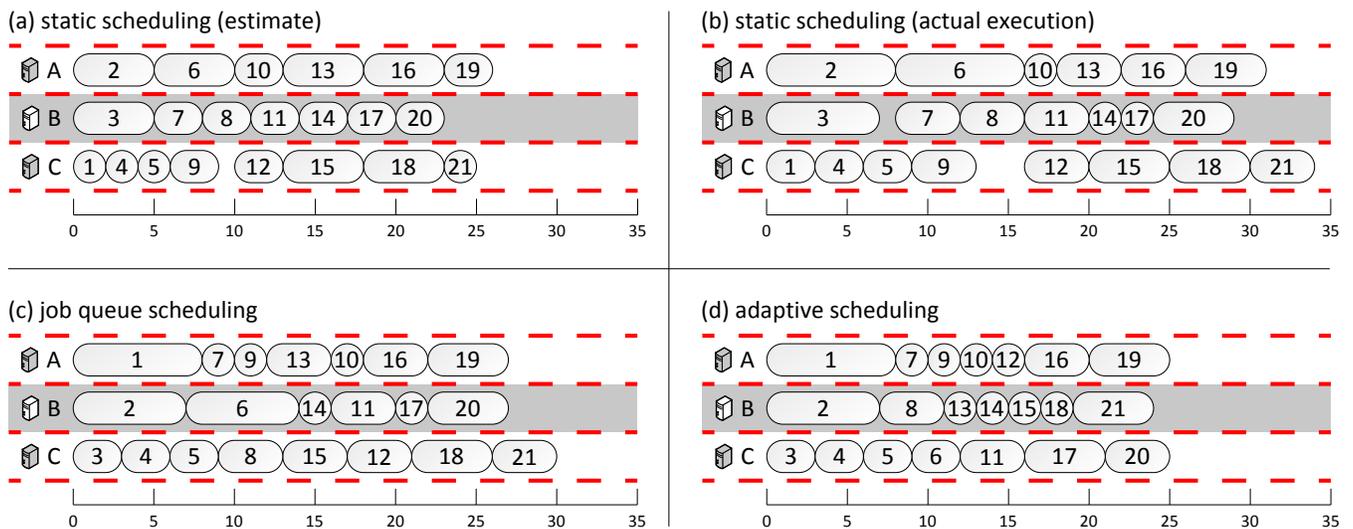}
  \caption[...]{Different scheduling techniques and their effect on the execution trace for the workflow from Figure~\ref{AlignmentWorkflowDataScheduling}. Horizontal lanes correspond to the compute resources A, B, and C. The numbers inside the task nodes match the numbers of tasks introduced in Figure~\ref{AlignmentWorkflowDataScheduling}. \textbf{(a)} A schedule obtained via static scheduling using the HEFT heuristic without incorporating data transfer times. \textbf{(b)} The execution trace generated when the HEFT schedule is strictly abided. \textbf{(c)} The execution trace if scheduling is performed using a simple job queue. \textbf{(c)} The execution trace for a form of adaptive scheduling, in which tasks are scheduled on resources on which they have performed above-average in the past.}
  \label{AlignmentWorkflowDataSchedules}
\end{figure*}

In static scheduling, schedules are assembled prior to workflow execution and strictly abided at runtime. While this method can yield good results in controllable or homogeneous compute environments, variations in resource performance can strongly impair overall execution time. Examples for \textit{static} scheduling strategies have been presented in~\citep{Yu06,Mandal2005,Blythe2005}. Pegasus is an example for a SWfMS implementing a static scheduling scheme~\citep{Deelman05}.

Figure~\ref{AlignmentWorkflowDataSchedules} (a) illustrates a static schedule assembled according to the Heterogeneous Earliest Finishing Time (HEFT) scheduling heuristic~\citep{Topcuoglu2002}. In HEFT, the workflow graph is traversed from the end to the beginning. Estimated times to finish workflow execution are computed at each task node, taking into account expected times for task execution and data transfer. To minimize overall time to completion, tasks with highest expected time to overall workflow completion are mapped onto faster resources. Replacing runtime estimates with actual task execution times, as listed in Figure~\ref{AlignmentWorkflowDataScheduling}, results in the execution trace illustrated in Figure~\ref{AlignmentWorkflowDataSchedules} (b) with an overall execution time of 34. Note that data transfer times are not considered in this example.

\subsubsection*{Job queue scheduling}
Job queue scheduling encompasses methods which assign tasks among compute resources in first-come-first-serve manner at runtime. The scheduler is oblivious to both the runtime statistics of the distributed compute infrastructure as well as the characteristics and requirements of individual workflow tasks. Implementations of job queue scheduling in SWfMS can be found in~\citep{Isard07,Missier2008,Warneke09}.

Figure~\ref{AlignmentWorkflowDataSchedules} (c) shows the execution trace for greedy job queue scheduling of the workflow from Figure~\ref{AlignmentWorkflowDataScheduling}. Here, tasks are put into a FIFO queue as soon as they're ready to execute. Idle compute resources extract tasks from this queue. This results in a total execution time of 30.

\subsubsection*{Adaptive scheduling}
An adaptive or dynamic scheduler actively monitors the computational infrastructure. It then adjusts workflow execution at runtime according to observed changes either by re-scheduling a previously assembled schedule (as proposed by~\citep{Lee2009,Sakellariou2004,Yu2007}) or by suspending resource assignment until a task is ready to execute (as implemented in DAGMan and Swift~\citep{Couvares2007,Wilde2011}).

A fully adaptive scheduler maps tasks onto suitable compute resources at runtime according to not only the current performance statistics of the resource, but also the specific requirements and characteristics of the task. For instance, a fully adaptive scheduler would attempt to assign an I/O-intensive task to a compute resource with above-average I/O-throughput. Clearly, reorganization of task execution at runtime comes at a cost, so there is a tradeoff to consider.

The possible benefits of fully adaptive scheduling are illustrated in Figure~\ref{AlignmentWorkflowDataSchedules} (d). Each compute resource keeps track of its execution times for certain tasks and compares their performance to other resources. Whenever it finishes the execution of a task, it chooses to execute a new task for which it knows to exhibit above average performance or for which it has not obtained performance statistics yet. This way, the workflow introduced in Figure~\ref{AlignmentWorkflowDataScheduling} could be executed in mere 25 units of time.

\section{Parallelism in current SWfMS}
\label{Systems}
\begin{figure}[t]
  \centering
  \includegraphics[width=\columnwidth]{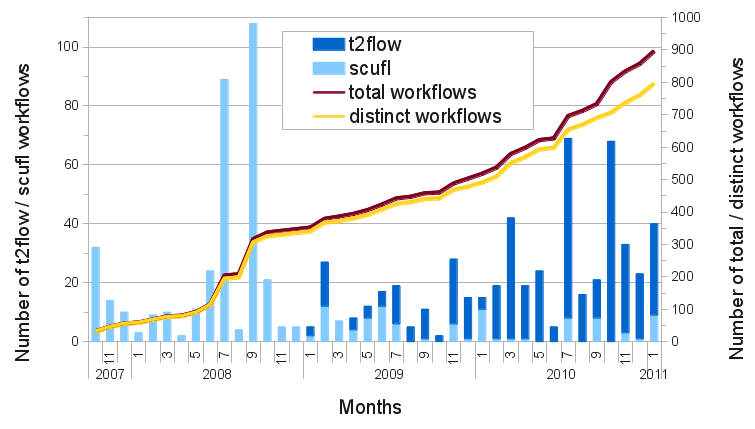}
  \caption[...]{Amount of Taverna workflows in Scufl and T2flow format uploaded to myExperiment per month as well as the total number of Taverna workflows (overall and distinct) in myExperiment (image taken from~\citep{Starlinger12}).}
  \label{myExperimentGrowth}
\end{figure}

In recent years, a number of concrete SWfMS have been developed and researchers have begun to adopt scientific workflows as model of computation of choice for their analysis pipelines. This development is reflected by the growth of public workflow repositories like myExperiment~\citep{Goble07} (see Figure~\ref{myExperimentGrowth}). At the same time, ever-increasing quantities of data generated in scientific experiments have elevated the demand for parallel execution of scientific workflows. While growing numbers of cores on servers (see Figure~\ref{Hasso}) along with novel computational infrastructures implementing sharing and leasing of compute resources can provide the computational backbone for massively parallel computation, scientific workflow enactment in parallel and/or distributed infrastructures brings with it a number of design choices, which were outlined in Section~\ref{Concepts}.

In this section, we give an overview of concrete support for parallelism in existing SWfMS. We focus our analysis to SWfMS that provide at least some minimal capabilities for parallel execution and that were actively maintained at the time of writing. These criteria apply well (but not exclusively) to the systems Swift, Condor DAGMan, Pegasus, Taverna, KNIME, and Kepler. While not technically SWfMS, we also shortly outline massively data parallel programming models like MapReduce and PACTs. We illustrate how these systems implement the concepts of parallelism outlined in Section~\ref{Concepts}, namely their supported types of parallelism, adoption of parallel computation infrastructures, and abilities with regard to scheduling. Table~\ref{parallelism} gives a summary of our findings.

\begin{table*}[t]
	\caption[Categorization of SWfMS]{Categorization of SWfMS}
	\label{parallelism}
	\medskip
	\centering
	\begin{tabular}{lllllll}
		\toprule
		SWfMS&explicitly supported&supported infrastructures&scheduling&refs\\
		&types of parallelism&&&\\
		\midrule
		Swift&task&multithreading, grid, cloud&adaptive&\citep{Zhao07,Wilde2011}\\
		Condor DAGMan&task&grid&adaptive&\citep{Litzkow88,Thain2005,Couvares2007}\\
		Pegasus&task&grid, cloud&static&\citep{Deelman05}\\
		Taverna&task, data, pipeline&multithreading&job queue&\citep{Oinn2004,Missier2008}\\
		KNIME&task, data&multithreading&job queue&\citep{Berthold2006,Sieb2007}\\
		Kepler&task, pipeline&multithreading&static, job queue&\citep{Ludascher06,Kepler2011}\\
		\bottomrule
	\end{tabular}
	\par
	\medskip
	\footnotesize
	A categorization of SWfMS with regard to their supported types of parallelism, computational infrastructure, and scheduling techniques. Multithreading capabilities indicate support for parallel execution on both local multicore machines and clusters.
\end{table*}

\subsection{Textual workflow languages}
\label{Textual}

This class of SWfMS encompasses textual languages which facilitate the enactment of large workflows in distributed compute infrastructures. In contrast to batch scripts, these languages provide support for highly parallel computation. As they were mostly designed to run on heterogeneous, geographically distributed compute resources, considerable administration effort is required for installation and utilization. Hence, the focus of these systems lies less on ease of use and more on efficient computing of high workloads. Like in a batch script, workflows are specified in a text file (like XML) according to a proprietary notation defined by the SWfMS.

\subsubsection{Swift}
Among the pioneers of parallel workflow execution is the Swift parallel scripting language~\citep{Zhao07}. Swift provides a functional language in which workflows are modeled as a set of program invocations with their associated command-line arguments as well as input and output files. Swift scripts are oblivious to the runtime environment and can be executed on local multicore computers, clusters, grids, or clouds. The execution engine of Swift supports task parallelism, as a task is dispatched for execution in a distributed environment as soon as all its input parameters are available. As data dependencies are resolved, the workflow is expanded dynamically at runtime.

For each known compute resource, the Swift execution engine maintains a score which increases with each successful task execution and decreases with each failed attempt~\citep{Wilde2011}. Tasks are assigned reactively to compute resources at runtime and the higher the score of a resource the more tasks will be assigned to it. Hence, changes in the performance of compute nodes are reflected in the score and taken into account by the scheduler. While scheduling reacts to alterations in the workload of a resource, it can't utilize elasticity in the form of provisioning of additional compute resources at runtime. Swift implements capabilities for failure recovery (retry, restart, and replication) as well as provenance tracking.

In summary, Swift implements adaptive scheduling and task parallelism on arbitrary compositions of local machines and clusters as well as grid and cloud infrastructures, as shown in Table~\ref{parallelism}. Swift has been utilized for computationally intensive applications from various fields of science, including physics and the life sciences. For an overview of computationally intensive scientific applications implemented in a SWfMS discussed in this survey, see Table~\ref{applications}.

\begin{table*}[hbtp]
	\caption[Scientific applications implemented SWfMS]{Scientific applications implemented in a SWfMS}
	\label{applications}
	\medskip
	\centering
	\begin{tabular}{llllll}
		\toprule
		SWfMS&life sciences&physics&astronomy&geography\\
		\midrule
		Swift&\citep{Stef-Praun2007,Hasson2008,DeBartolo2010,Lee2010,Adhikari2012}&&&\citep{Agarwal2011,Woitaszek2011}\\
		Pegasus&\citep{Wang2011,Mehta2011}&\citep{Brown2006}&\citep{Berriman2004}&\citep{Deelman2006}\\
		Taverna&\citep{Kell2009,Maleki-Dizaji2009,Zhou2009}&&&\\
		Kepler&\citep{Michener2007}&\citep{Podhorszki2007}&&\citep{Barseghian2010}\\
		\bottomrule
	\end{tabular}
\end{table*}

\subsubsection{Condor DAGMan}
Condor~\citep{Litzkow88} is a batch job scheduler for high-throughput computing on distributed resources. It was originally designed to scavenge idle workstations for CPU cycles, yet has been extended with functionality to interface grid resources via Condor-G. Condor puts a strong emphasis on reliability of execution in the form of job checkpointing, recovery, and migration. 
Condor provides a resource allocation language called \textit{ClassAds}~\citep{Thain2005} (``classified advertisements'') to describe requirements and preferences for matchings between tasks and compute nodes. For instance, ClassAds can be utilized to describe that a machine only accepts tasks if its current workload is low and it has been idle for an hour or that a task prefers to be executed on a machine with good floating point performance. Condor maintains a queue of tasks in which new tasks are scheduled on the compute resource with the best matching ClassAd. Scheduling in Condor can therefore be considered adaptive, provided the ClassAds are specified such that machines with a high workload are less likely to be assigned new tasks.

Condor's Directed Acyclic Graph Manager (DAGMan)~\citep{Couvares2007} provides the means to textually specify a workflow as a DAG describing a set of tasks along with their data interdependencies. DAGMan operates as a Condor job and supervises workflow execution, submitting workflow tasks which are ready for execution to Condor one at a time. Data parallelism and pipelining are not natively supported since a task is not submitted for execution until all of its input data is available. DAGMan does not communicate with the Condor scheduler, hence advanced workflow characteristics such as task runtime estimates are not considered in scheduling. Since DAGMan does not manage the movement of intermediate data products between jobs, data movements have to be explicitly specified within the workflow DAG. Similar to Condor, DAGMan emphasizes reliability. In case of failure, DAGMan compiles a rescue DAG from which execution can be resumed.

As shown in Table~\ref{parallelism}, the SWfMS DAGMan supports adaptive scheduling and parallel task execution on a grid infrastructure. Since a compute grid might not be available to some users, efforts have been made to execute DAGMan workflows on local compute infrastructure without Condor (e.g., \citep{Schmidt2012}).

\subsubsection{Pegasus}
DAGMan does not provide the means to automatically set up auxiliary tasks, such as data movement, cleanup, or workflow optimization. Therefore, \citet{Deelman05} developed Pegasus as a data-aware layer on top of DAGMan, which introduces capabilities for provenance tracking, execution monitoring, and failure recovery. The SWfMS Pegasus can be viewed as a collection of DAG transformers which iteratively translate a concrete workflow into an executable DAGMan workflow. To this end, Pegasus queries and maintains catalogs of available computational resources, data replicates, and data processing software/services. Pegasus also prunes and optimizes workflow structure by omitting tasks for which output data has been computed previously and by clustering short-running tasks into joint DAGMan jobs.

Tasks in the DAGMan input file generated by Pegasus are location-specific, i.e.,~Pegasus overrides the default scheduling mechanism of Condor (\textit{ClassAds}). Instead, Pegasus provides four different scheduling strategies by default:

\begin{itemize}
	\item \textit{Random}: Tasks are randomly assigned to compute resources able to execute them.
	\item \textit{Round-Robin}: Tasks are evenly distributed among resources, independent of the associated computational cost.
	\item \textit{Group}: Tasks can be put into user-defined groups. Each group of tasks is scheduled to run on the same compute resource.
	\item \textit{HEFT}: The Heterogeneous Earliest Finishing Time scheduling heuristic~\citep{Topcuoglu2002} described in Section~\ref{Scheduling}. Pegasus assumes default costs for data communication, whereas the runtime for tasks has to be specified by the user.
\end{itemize}

All these scheduling strategies result in a static schedule. Since the workload on the compute infrastructure might be subject to change, \citet{Lee2009} developed an adaptive scheduling mechanism for Pegasus. Here, job queues on execution sites are observed and compared. In case of sustained discrepancies, Pegasus re-schedules the workflow from the current point of execution.

Pegasus was originally designed to distribute computationally intensive workflows across grid infrastructures. However, the growing interest in cloud computing has led to efforts to efficiently run Pegasus in a cloud infrastructure~\citep{Hoffa08,Juve09,Juve11}. In its productive version, Pegasus supports task parallel execution of scientific workflows on grid and cloud infrastructures using a static schedule (see Table~\ref{parallelism}). Several computationally intensive workflows from different scientific domains have been implemented and executed in Pegasus (see Table~\ref{applications}).

\subsubsection{Other textual workflow languages}
While aforementioned textual workflow languages can manage execution of thousands of tasks in parallel, evaluation of workflow files (e.g., Swift scripts) is performed on a single node. In the light of ever-increasing concurrency in today's computing systems, this may constitute a bottleneck for workflows consisting of very large numbers of short-running tasks. For this reason, \citet{Wozniak2012} implemented Turbine, an extreme-scale workflow management system with an execution engine distributed over multiple compute nodes, which is responsible for resolving data dependencies, managing task distribution and load balancing, and providing global data storage. Turbine can compile Swift scripts and has been shown to be able to distribute more than 20,000 workflow tasks per second.

\citet{Islam2012} found many established workflow management systems to lack in scalability. Hence, they developed Oozie, a scalable workflow scheduler which is built on top of Hadoop~\citep{White10}, the open source implementation of the MapReduce~\citep{Dean08} programming paradigm. The Oozie server accepts textually specified workflow DAGs submitted by multiple Oozie clients, 
splits these workflows into sub-tasks, and dispatches the sub-tasks to a Hadoop cluster for processing. Oozie is highly scalable and has been utilized by Yahoo! for execution of more than 770,000 workflows.

\citet{Ogasawara11} recently presented a relational algebra that facilitates highly scalable scientific workflow execution. In their work, data is represented in the form of relations. Workflow tasks, such as external program or service invocations, are interpreted as one of four operators. These operators -- \textit{Map}, \textit{SplitMap}, \textit{Reduce}, and \textit{Filter} -- are characterized by the ratio at which they consume and produce tuples of input and output data. Workflows modeled as compositions of these operators can be structurally optimized by applying concepts of database query optimization, such as predicate pushdown. The relational view on data enables the scheduler to automatically distribute fragments of the workflow and parts of the data among several processing nodes, resulting in a task and data parallel execution. Different scheduling strategies have been shown to yield runtime gains, depending on workflow structure and its opportunities for structural optimization.

\begin{figure*}[t]
  \centering
  \includegraphics[width=\textwidth]{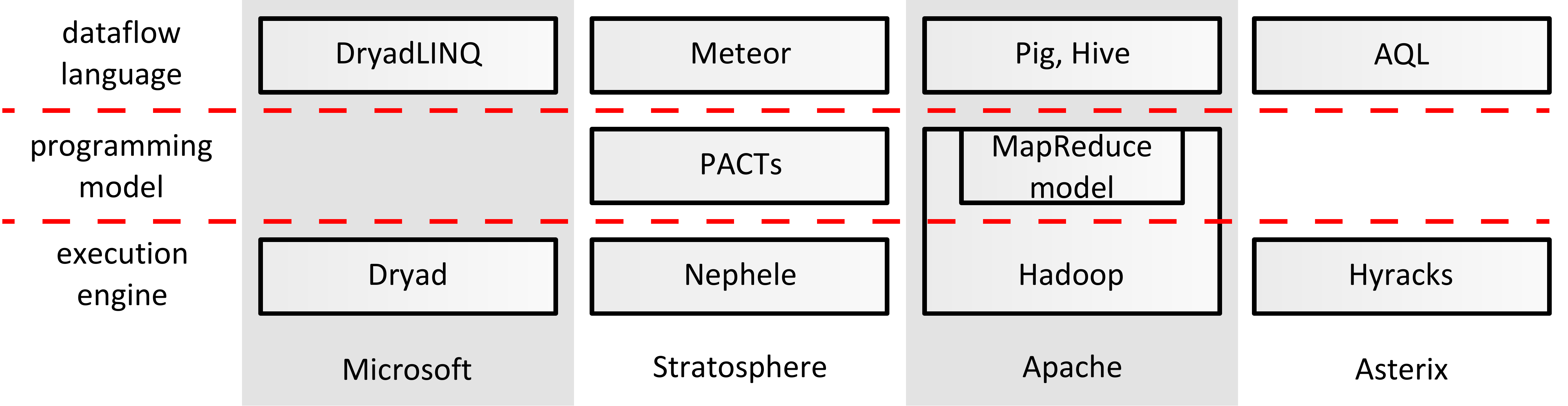}
  \caption[...]{Stack architecture of the dataflow languages DryadLINQ, Meteor, Pig, Hive, and AQL.}
  \label{dataflow}
\end{figure*}

Instead of designing workflows \textit{ad hoc} in a way that allows for data parallel execution, \citet{DeOliveira2010} proposed to outsource only the most computationally taxing workflow tasks to external resources. Following this rationale, they developed \mbox{SciCumulus}, which can be described as a cloud-based scientific workflow middleware. Scientists can wrap computationally intensive tasks of their existing workflows into \mbox{SciCumulus} cloud activities. To this end, \mbox{SciCumulus} provides components for upload, dispatch, download, and provenance capture. These components can be interfaced from within a SWfMS like Taverna, Pegasus, or Kepler. By using predefined or custom cartridges, users can specify how data is to be fragmented before and merged after processing for data parallel computation. \mbox{SciCumulus} can also be employed to perform parameter sweeps on workflow components in parallel.

\citet{Cieslik11} developed PaPy (``Parallel pipelines in Python''), a light-weight and modular textual SWfMS in which workflow tasks are specified as Python functions. PaPy supports task parallelism in the form of a worker pool distributed among local and remote resources. Furthermore, by allowing input data to be split and processed in chunks, PaPy can be configured to perform data parallel computation. 


\subsubsection{Massively data parallel query languages}
Subsequent to the publication and wide-spread adoption of Google's MapReduce~\citep{Dean08} programming model and its open source implementation Hadoop~\citep{White10}, a class of systems that can be summarized under the term ``dataflow languages'' have emerged. These textual languages including DryadLINQ~\citep{Yu08} from Microsoft, Stratosphere's Meteor~\citep{Heise2012}, Pig~\citep{Olston08} from Yahoo!, Hive~\citep{Thusoo2009} from Facebook, and the Asterix Query Language (AQL)~\citep{Behm2011} from UC Irvine were designed to efficiently perform query-style dataflow programs over extremely large data in parallel.

Dataflow programs specified in these languages are translated into DAGs of Dryad vertices, Stratosphere parallelization contracts (PACTs~\citep{Battre10}), Hadoop \textit{map} and \textit{reduce} jobs, or Hyracks~\citep{Borkar2011} operators. Similar to scientific workflows, these DAGs describe data dependencies between separate data processing steps. Data parallel execution of these DAGs is handled by the execution engines underlying the Dryad, Hadoop, and Hyracks implementation or -- in the case of Stratosphere -- the Nephele~\citep{Warneke09} scheduler. See Figure~\ref{dataflow} for a graphical overview of the stack architecture of aforementioned systems.


While all of these dataflow languages provide the means to model computationally intensive problems, they were not primarily designed for scientific data. Thus, in contrast to most SWfMS, they neither provide domain-specific software libraries nor are there any published workflows designed by domain scientists. Furthermore, while each task in a scientific workflow is treated as a black box, dataflow languages often require the user to specify the task according to a restrictive query-based syntax, which is typically not easily accessible by non-computer-savvy users. For these reasons, while MapReduce and similar frameworks have an undeniable influence on parallelization in SWfMS, they are out of scope for this survey.

\looseness=-1However, Dryad~\citep{Isard07} and Nephele~\citep{Warneke09}, the execution engines underlying the dataflow languages DryadLINQ and Meteor, respectively, can both be utilized for execution of arbitrary workflows. Similar to DAGMan or Pegasus, Dryad maps abstract workflows specified as DAGs to their concrete compute resources. However, while in Pegasus tasks can only exchange data in the form of files, Dryad also supports network or shared memory communication. These features enable Dryad to perform a pipeline parallel execution in which tasks start execution as soon as at least one input record is available at each input port. Dryad tasks are written in C++ and tasks can be annotated with preferences or constraints towards the resources on which they are to be executed. This puts data-awareness into the hands of the user, since the default scheduling strategy of Dryad is a job queue in which new tasks are greedily assigned to the first available resource. 

The Dryad framework was developed with a static computation environments in mind. In contrast, Nephele~\citep{Warneke09} also supports distribution of tasks on dynamically scalable compute infrastructures, such as compute clouds. Similar to Dryad, Nephele supports task communication in the form of shared memory and TCP network connections. Users can also specify to cluster several short-running tasks on a single compute resource or split long-running tasks into subtasks for data parallel execution. If no annotations are provided by the user, the Nephele scheduler assigns each task to its own compute resource. 

In summary, 
Dryad and Nephele support task, data
, and pipeline parallelism on multicore architectures and compute clouds using a job queue for just-in-time scheduling.

\subsection{Graphical workflow management systems}
\label{Standalone}
The class of graphical SWfMS comprises systems with a strong emphasis on ease of use and graphical representation of workflows. Since graphical representation becomes problematic for large workflows consisting of hundreds or thousands of tasks~\citep{Deelman09}, most graphical SWfMS provide a means to compose workflows from a hierarchical nesting of subworkflows. In contrast to textual workflow scripting languages, workflows are mostly designed from collections of web-services or built-in components which can be adjusted by various parameters.

Graphical SWfMS are typically installed either on a remote server or on a local client and accessed using a graphical user interface with drag-and-drop functionality on the client. Hence, workflow execution is performed only on a single machine. 

Graphical systems sometimes support multithreading, but most of them can't utilize external compute resources by default, save via integration of web services. However, approaches towards execution on distributed computational infrastructures are increasingly being explored. For scheduling, tasks with available input data are usually put into a job queue and executed by replicate workers from a thread pool.

\subsubsection{Taverna}
Taverna Workbench~\citep{Oinn2004} is an established graphical SWfMS developed for the enactment of bioinformatics workflows. It emphasizes usability, providing a graphical user interface for workflow modeling and monitoring as well as a comprehensive collection of pre-defined services. Taverna workflows are internally represented in one out of the two textual languages Scufl (used by Taverna 1) and T2flow (used by Taverna 2). User-generated workflows can be exchanged through the myExperiment workflow repository~\citep{Goble07}. See Figure~\ref{TavernaWorkflow1172} for a bioinformatics Taverna workflow from myExperiment.

\looseness=1Taverna workflows can be executed either on the client or on a server. As of yet, Taverna can utilize only a single machine, as there is no support for execution of computationally intensive tasks on more than one server or on distributed architectures, such as grid or cloud infrastructures. 

In Taverna, a task can start as soon as input data is available at all of its input ports. Each task is processed by a separate thread, with the maximum number of concurrently running threads being set by the user. If a task receives a list of data items on an input port where a single item is expected, each element of the list is processed by a replicate of the task in a separate thread~\citep{Missier2008}. Each processed data item is passed to following tasks for immediate consumption in a new thread. Taverna's implicitly pipeline parallel execution model results in multiple replicate data processing pipelines running concurrently.

In summary, while Taverna provides strong support towards parallelized workflow enactment in the form of task, data, and pipeline parallelism, it is devoid of more sophisticated scheduling methods than a job queue and can currently only utilize cores on a single local resource (see Table~\ref{parallelism}). Several computationally intensive problems from the field of bioinformatics have been implemented in Taverna, as shown in Table~\ref{applications}.

\subsubsection{KNIME}
The Konstanz Information Miner (KNIME)~\citep{Berthold2006} shares many characteristics with Taverna, albeit with a stronger focus on user interaction and visualization of results, yet with a smaller emphasis on web service invocation. Furthermore, KNIME focuses on workflows from the fields of data mining, machine learning, and chemistry, while Taverna is more concerned with integration of distributed and possibly heterogeneous data. A graphical user interface facilitates design and execution monitoring of workflows. KNIME can either be installed locally or on a server, accessible via multiple clients.

\looseness=1As in Taverna, each task ready for execution is put into a queue, from which it is retrieved and processed by a separate thread. The size of the thread pool is restricted via user-defined constraints. While Taverna detects opportunities for data parallelism by monitoring the input and output ports of a task, KNIME requires the designer of a task node to explicitly specify whether it qualifies for data parallel execution. If implemented accordingly, KNIME automatically splits the entire input data into four times as many chunks as the size of the thread pool. Similar to data parallelism in \textit{map}-tasks of the MapReduce paradigm, each chunk is then processed by a replicate task and aggregate results are merged as soon as all threads have finished execution~\citep{Sieb2007}. Note that in contrast to Taverna, KNIME does not support pipeline parallelism, as a task can't start execution until all of its parent tasks have processed all of their data.

As shown in Table~\ref{parallelism}, KNIME implements job queue scheduling as well as task and data parallelism in the form of multiple threads on a local machine or on a remote server. Distributed architectures like a compute grid or cloud are not supported by default.

\subsubsection{Kepler}
\begin{figure}[t]
\begin{center}
  \includegraphics[width=\columnwidth]{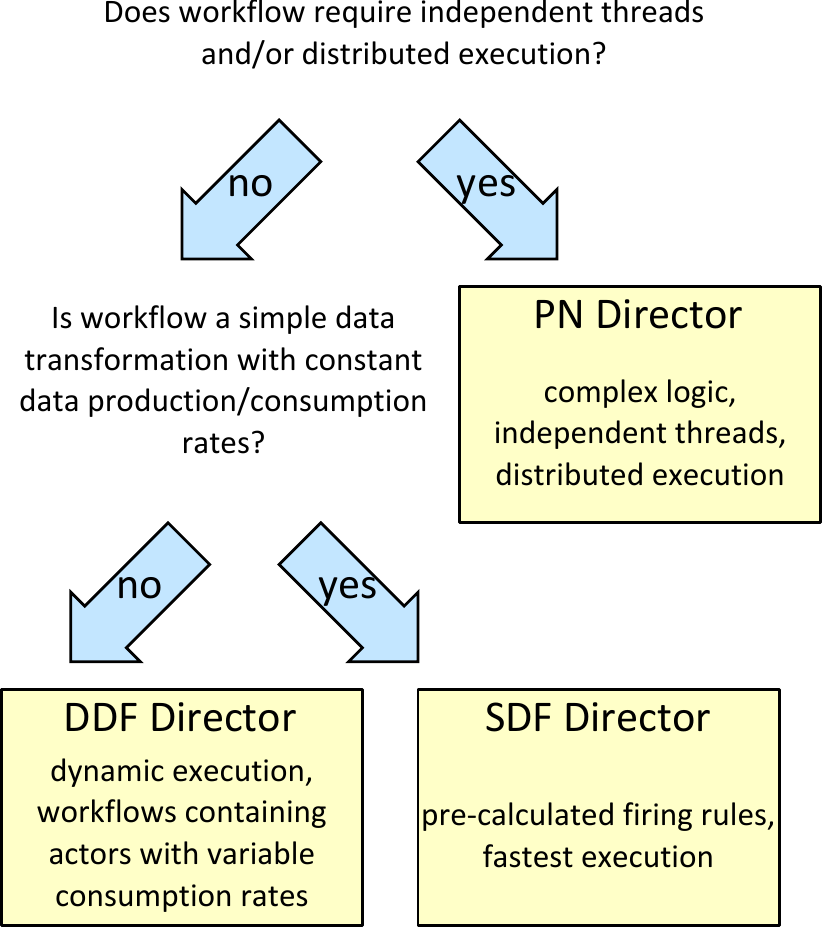}
\end{center}
  \caption{Guideline for choosing a director based on a given Kepler workflow~\citep{Kepler2011}.}
  \label{KeplerDirector}
\end{figure}

Kepler~\citep{Ludascher06} is a frequently used graphical SWfMS. Similar to Taverna and KNIME, it provides an assortment of built-in components with a major focus on statistical analysis. Neither a server-based implementation of Kepler nor support for distributed execution are currently provided. Kepler is built on top of the Ptolemy II Java library~\citep{Davis2001}, from which it inherits the concept of the so-called \textit{director}. By choosing a director, users can specify how a workflow is processed, i.e.,~which scheduling technique should be employed. See Figure~\ref{KeplerDirector} for a guideline on which director to choose for a given Kepler workflow. Default options include the synchronous dataflow director (SDF), dynamic dataflow director (DDF), and process network director (PN)~\citep{Kepler2011}.

\begin{table*}[t]
	\caption[Comparison of graphical SWfMS]{Comparison of graphical SWfMS}
	\label{comparison}
	\medskip
	\centering
	\begin{tabular}{llll}
		\toprule
		&major foci&strongly supported domains&data parallelism, pipelining\\
		\midrule
		Taverna&web service composition&life sciences&enabled by default\\
		KNIME&visualization, user interaction&chemistry, machine learning&only selected components\\
		Kepler&control over dataflow (director)&statistics&if PN director is selected\\
		\bottomrule
	\end{tabular}
\end{table*}

SDF and DDF directors execute the whole workflow in a single thread. In contrast, the PN director assigns a separate thread to each workflow task, implementing a pipelined execution on local Java threads. Data items are passed from the output port of parent tasks to the input port of child tasks, where they are stored in a buffer. The PN director supervises workflow execution by monitoring buffer sizes and instructing threads to process new data items when available. The choice of the director also defines whether scheduling be performed statically and prior to execution (SDF director) or at runtime (DDF and PN director).

In an attempt to enable data parallel execution of data-intensive Kepler workflows,~\citet{Wang09} presented \textit{map} and \textit{reduce} actors based on the parallel computing framework Hadoop. By integrating the programming model of MapReduce into Kepler, workflow designers can benefit from the parallel programming model of MapReduce without having to worry about the programming interfaces. However, a generic word count problem implemented in Kepler using this MapReduce component showed clearly inferior performance when compared to the equivalent Java Hadoop implementation~\citep{Wang09}. This is most likely the consequence of the Kepler engine having to be initiated separately and repeatedly for each of the \textit{map} and \textit{reduce} tasks.

To summarize, by default Kepler supports task and pipeline parallel execution of workflows in possibly multiple threads on a local client (see Table~\ref{parallelism}). Scheduling is performed statically in advance or just-in-time using a job queue. Several reports of scientific projects using the Kepler SWfMS have been published, as shown in Table~\ref{applications}.

A guideline on when to choose which of the SWfMS Taverna, KNIME, and Kepler is given in Table~\ref{comparison} in form of a comparison.

\subsection{Domain-specific web portals}
\label{CloudLife}
This class of SWfMS is comprised of systems that facilitate composition of domain-specific data repositories, web services, and applications. A strong emphasis is usually put on easy set up, intuitive operation, and sharing of workflows. In most cases, workflows are designed in a web browser and executed on a public or private server. In this survey, we limit our examination of domain-specific web portals to SWfMS specific to the life science. The rationale behind this is that many SWfMS have been tailored specifically to the life sciences and a considerable number of computational tasks in bioinformatics qualify for parallel execution.

With the advent of next generation sequencing and the general exponential rise of data in bioinformatics, Galaxy \citep{Goecks10} has been established as one of the major frameworks for genomic research in the life sciences. Galaxy comes with a web-based user interface as well as assorted pre-built components for popular tasks in sequence analysis. Workflows can be assembled from tasks and data repositories, shared with other users and executed on a public or private server. See Figure~\ref{GalaxyWorkflow} for a Galaxy workflow from the public Galaxy server. \citet{Afgan2010} recently presented CloudMan as a scalable resource system for a deployment of Galaxy servers in an EC2 Cloud. Note that CloudMan does not distribute tasks or data of an individual workflow among virtual machines. It merely allows multiple workflows to be run on different virtual machines.

The lack of interoperability between the two major SWfMS in the field of bioinformatics, Galaxy and Taverna, severely hampers workflow and knowledge sharing within the scientific community. Therefore, \citet{Abouelhoda2012} developed Tavaxy, a stand-alone SWfMS which can integrate both Taverna and Galaxy workflows at design-time as well as at run-time. Tavaxy workflows are specified in tScufl, which is highly inspired by Taverna's Scufl language, whereas the workflow execution engine underlying Tavaxy can be considered an extension of the Galaxy engine. Publicly available VM images of Tavaxy provide a means to execute workflow tasks, subworkflows or entire workflows on rented Amazon EC2 cloud infrastructure.

\citet{Linke11} observed that most bioinformatics workflows designed in Taverna and Galaxy are dedicated mostly to data conversion, which often results in the loss of complementary data. Furthermore, with high-throughput experiments producing continuously more data, processing via remote web services is often not feasible. Therefore, they presented Conveyor as a client/server-based bioinformatics workflow engine with a strongly typed hierarchy of data types and an increased focus on local execution. All tasks composing a Conveyor workflow are executed in a separate thread, effectuating in task parallel execution.

The typical biologist interested in a bioinformatics analysis often has to consult several online data repositories and combine various analytic tools manually. To facilitate this process, \citet{Neron09} developed Mobyle, a framework accessed from within a web browser that federates local and remote bioinformatics services and simplifies their composition. Administrators can register new services on any Mobyle server and local clients can access the combined services provided by several of these servers. Parallel execution of tasks on different servers is handled by Mobyle's internal workflow engine.

In bioinformatics, progress in high-throughput technologies demands continuous development of novel computational analysis methods. \citet{Kallio2011} found the usage of most of these methods to require considerable computational skill. Hence, they developed Chipster, an easy-to-use workflow management system featuring a large collection of built-in data analysis methods as well as interfaces to add newly developed methods. Workflow design is performed in a graphical user interface on a locally installed client, whereas most of the computation is conducted on a server. The server can be set up to incorporate analysis tools spread across several computational nodes, employing the concept of task parallelism.

Several computationally intensive analysis methods in bioinformatics have recently been implemented using the MapReduce programming paradigm (e.g., sequence alignment using CloudBurst~\citep{Schatz2009}). Since installing and utilizing MapReduce-based applications on a distributed computational infrastructure might be difficult for domain scientists, \citet{Schoenherr2012} developed Cloudgene as an extensible execution environment for MapReduce programs in bioinformatics featuring a graphical user interface and a selection of built-in components. Cloudgene allows the graphical design and distributed execution of analysis pipelines on local clusters or public clouds, such as Amazon EC2.

Another example of employing the MapReduce programming paradigm for bioinformatics workflows was presented by \citet{Wu11}. They developed a life science gateway that allows for data parallel execution of workflows in the cloud based on Hadoop streaming. Their framework lets users specify embarrassingly parallel workflows, such as pairwise BLAST sequence alignment, as a series of map and reduce tasks, with each task being specified in command-line syntax. Scientists using this framework can also share data and workflows, similar to Galaxy and myExperiment.

In the light of overwhelming assortments of bioinformatics applications and libraries, users might prefer to employ pre-built workflows recommended by experts instead of assembling their own data processing pipelines. Following this rationale, \citet{Angiuoli11} developed the Cloud Virtual Resource (CloVR), a life science gateway featuring a selection of four hard-coded workflows covering some of the major tasks in next generation sequence analysis. These workflows are encased by a virtual machine image and are therefore easy to set up and execute. Computationally taxing BLAST~\citep{Altschul1997} sequence searches occurring in three of the workflows are split and distributed among dynamically extendable cloud resources, according to a BLAST runtime estimation. While the CloVR virtual machine images provide additional pre-installed frameworks for parallel execution, such as Hadoop, they are not utilized by any of the default workflows.

\section{Conclusion and Discussion}
\label{Discussion}
Scientific workflows have recently emerged as a model of computation for processing of scientific data. However, increasing amounts of data as well as a growing interest of the scientific community in data-driven research have eventuated in increasingly high requirements of computational power and thus a growing demand for parallelization techniques.

\looseness=1In Section~\ref{Concepts}, we outlined three major aspects of parallelism in SWfMS: types of parallelism, distributed compute infrastructures, and approaches towards scheduling. With regards to these concepts, we observed three classes of established SWfMS in Section~\ref{Systems}: (1) textual workflow languages, which can distribute workflow tasks over external resources, but are difficult to set up by domain scientists and often lack support for data and pipeline parallelism (Swift, DAGMan, Pegasus); (2) graphical standalone systems, which are easy to use but are not able to efficiently integrate external resources (Taverna, KNIME, Kepler); (3) life science enactment portals in which domain scientists can design workflows in a web browser as well as execute and share their workflows on a remote and possibly public server (Galaxy, Mobyle, Conveyor). We argue that all of these approaches leave considerable room for improvement:

\begin{itemize}
	\item \textit{Types of parallelism}: While textual workflow languages provide the capabilities to distribute workflow tasks over external compute resources, they often don't support data or pipeline parallelism. Despite not all workflows qualifying for data parallel execution, many of them do and hence much potential for highly scalable scientific workflow enactment remains untapped in textual languages. At the same time, graphical standalone systems like Taverna implement data and pipeline parallelism yet can't schedule threads on external resources. We believe that the computational model of scientific workflows could benefit greatly if both classes of systems would complement or inherit concepts from one another.
	\item \textit{Parallel compute infrastructures}: Over the last few decades, the number of cores per CPU and CPUs per cluster has been continuously rising. In more recent years, cloud computing technology has reached the stage of productivity, providing highly scalable compute resources on demand. Renting computational platforms as a service seems especially beneficial for those research applications where high-performance compute resources are only required infrequently whenever new experimental data has been produced. While most SWfMS have been adapted to support multithreading on multicore architectures, few SWfMS are able to incorporate grid and cloud resources and none currently offer advanced features like autonomous (de-)allocation of machines at runtime. We argue that scientific workflow technology requires new models of execution inherent to distributed infrastructures in order to stay competitive.
	\item \textit{Scheduling}: The physical infrastructure underlying distributed compute environments like grids and clouds is often shared between multiple users. As users execute programs with different hardware requirements (e.g.,~I/O-bound vs.~CPU-bound), the performance of compute nodes can dynamically change at runtime. Scheduling of scientific workflows is therefore best approached adaptively, by finding suitable matches between workflow tasks and compute nodes based on statistics obtained at runtime. Although advanced scheduling techniques have been shown to net considerable runtime improvements, we observed that most SWfMS rely on very basic scheduling techniques such as greedy job queues or static \textit{a priori} assignments of tasks to compute nodes. Introducing support for multiple types of parallelism and computation on distributed infrastructures in SWfMS currently not supporting these concepts translates into more scheduling options and will therefore further leverage the importance of scheduling.
	\item \textit{Ease of use}: We observed that SWfMS either focus on usability (as seen in most graphical SWfMS and life science portals) or provide support for distributed computation (as found in most textual workflow languages). We argue that this gap between ease of use for non-computer savvy people and highly scalable execution has to be narrowed in order for the scientific community to adopt scientific workflows as model of computation of choice for their analysis pipelines.
	\item \textit{Structural workflow optimization}: Scientific workflows like Montage~\citep{Berriman2004} consist of a large number of short-running tasks. In distributed infrastructures, these short-running tasks can introduce considerable runtime overhead as a result of latency due to network connectivity and workflow engine initialization times. Automatic clustering of short-running tasks into composite tasks as a means of workflow optimization has been shown to yield promising results. Furthermore, many scientific workflows involve selection, filtering, and sorting steps on large collections of data, similar to database queries. Approaches towards structural workflow optimization inspired by established database query optimization techniques can constitute a valuable addition to parallel execution with the aim to reduce workflow runtime (e.g.,~\citep{Kougka2012}). Unfortunately, support for structural workflow optimization in concrete SWfMS is very limited and can be improved upon.
\end{itemize}

In closing, while substantial progress has been made in parallel scientific workflow enactment, the field of solutions is still heterogeneous and leaves room for improvement. The proliferation of cloud technologies will change the computational landscape of data-driven research. This will inevitably require SWfMS technology to adapt.

In particular, the following open research topics arise from the points discussed above:

\begin{enumerate}
	\item Extend current SWfMS with easy to set-up integration of local, grid, and cloud infrastructures as well as arbitrary compositions thereof.
	\item Investigate approaches towards adaptive scheduling in heterogeneous, dynamically changing computational environments, such as shared or composite infrastructures. 
	\item Refine SWfMS to be both easy to use for domain scientists and provide inherent data parallel execution for computationally expensive tasks on massive data.
	\item Explore new means of structural workflow optimization inspired by traditional database query optimization.
	\item Examine possibilities towards storing and searching workflows along with their execution traces in public repositories in order to reduce redundancy in design and execution of workflows.
\end{enumerate}


\section*{Acknowledgements}
We thank Astrid Rheinl\"{a}nder, Matthias Sax, Steffen Zeuch, and J\"{o}rgen Brandt for helpful comments on the manuscript.

\section*{Funding}
Marc Bux is funded by the Deutsche Forschungsgemeinschaft through graduate school SOAMED (GRK 1651).

\bibliographystyle{model1-num-names-custom}
\bibliography{library}


\end{document}